\documentclass[a4paper,12pt]{article}
\pdfoutput=1

\usepackage{amsfonts,amsthm,amsmath,amssymb,graphicx,hyperref,cite}
\usepackage[utf8]{inputenc}
\usepackage[all]{xy}
\usepackage[vcentermath]{youngtab}

\def\bea{\begin{eqnarray}}
\def\eea{\end{eqnarray}}

\def\a{\alpha}

\def\s{\sigma}

\def\r{\rho}
\def\b{\beta}
\def\m{\mu}
\def\n{\nu}

\begin{document}
\makeatletter
\@addtoreset{equation}{section}
\makeatother
\renewcommand{\theequation}{\thesection.\arabic{equation}}
\vspace{1.8truecm}

{\LARGE{ \centerline{\bf Emergent Yang-Mills theory}  }}

\vskip.5cm

\thispagestyle{empty}
\centerline{ {\large\bf
Robert de Mello Koch$^{a,b,}$\footnote{{\tt robert@neo.phys.wits.ac.za}},
Jia-Hui Huang$^{a,}$\footnote{{\tt huangjh@m.scnu.edu.cn}},
Minkyoo Kim$^{b,}$\footnote{{\tt minkyoo.kim@wits.ac.za}} }}\par
\vspace{.2cm}
\centerline{{\large\bf and
Hendrik J.R. Van Zyl${}^{b,}$\footnote{ {\tt hjrvanzyl@gmail.com}} }}

\vspace{.4cm}
\centerline{{\it ${}^{a}$ Guangdong Provincial Key Laboratory of Nuclear Science, Institute of Quantum Matter},}
\centerline{{ \it South China Normal University, Guangzhou 510006, China}}

\vspace{.4cm}
\centerline{{\it ${}^{b}$ National Institute for Theoretical Physics,}}
\centerline{{\it School of Physics and Mandelstam Institute for Theoretical Physics,}}
\centerline{{\it University of the Witwatersrand, Wits, 2050, } }
\centerline{{\it South Africa } }

\vspace{1truecm}

\thispagestyle{empty}

\centerline{\bf ABSTRACT}

\vskip.2cm
We study the spectrum of anomalous dimensions of operators dual to giant graviton branes.
The operators considered belong to the su$(2|3)$ sector of ${\cal N}=4$ super Yang-Mills theory, have a bare
dimension $\sim N$ and are a linear combination of restricted Schur polynomials with $p\sim O(1)$ long rows or columns.
In the same way that the operator mixing problem in the planar limit can be mapped to an integrable spin chain, we find
that our problem maps to particles hopping on a lattice.
The detailed form of the model is in precise agreement with the expected world volume dynamics of $p$ giant graviton branes,
which is a U$(p)$ Yang-Mills theory.
The lattice model we find has a number of noteworthy features.
It is a lattice model with all-to-all sites interactions and quenched disorder.

\setcounter{page}{0}
\setcounter{tocdepth}{2}
\newpage
\setcounter{footnote}{0}
\linespread{1.1}
\parskip 4pt

{}~
{}~

\section{Introduction}

The operator mixing problem in the planar limit of ${\cal N}=4$ super Yang-Mills theory is solved.
This dramatic progress was achieved by mapping the dilatation operator of the theory to the Hamiltonian
of an integrable spin chain\cite{Beisert:2010jr}.
The mapping identifies each single trace operator with a state of the spin chain and operators of a definite
dimension map to spin chain states with a definite energy.
The integrable model describes the dynamics of magnons which can scatter with each other.
This scattering between the magnons happens in one dimension.
As far as the single trace operators are concerned, reordering fields within the trace corresponds to changing their
positions in this single dimension.
In the integrable spin chain, this dimension is that of the spin chain lattice, while in the holographically dual theory
it is the string world sheet.
This is precisely what we should have expected from the AdS/CFT
correspondence\cite{Maldacena:1997re,Gubser:1998bc,Witten:1998qj}: we know that the planar limit of
the gauge theory is dual to perturbative string theory, so we expect the world sheet dynamics of a string to emerge
from the planar limit of the CFT.

We expect something similar happens whenever we focus on a class of operators that are holographically dual
to a system with a definite semi-classical limit: the dilatation operator should be mapped to the Hamiltonian of the
dynamics of the relevant semi-classical physics.
Our goal in this article is to test this expectation for the class of operators holographically dual to giant graviton
branes\cite{McGreevy:2000cw,Grisaru:2000zn,Hashimoto:2000zp}.
This class of operators have a bare dimension of order $N$\cite{CJR,David}.
In this regime the single trace operators don't provide a useful starting point for the operator mixing problem.
Indeed, for operators with such a large bare dimension mixing between different trace structures is not
suppressed\cite{Balasubramanian:2001nh}.
We will start from the basis provided by the restricted Schur polynomials, which is reviewed in Section \ref{bases}.
We study the su$(2|3)$ sector of the theory.
Truncation to this subsector is consistent to all orders of perturbation theory\cite{Beisert:2003jj}.
This is the maximal closed subsector with finitely many fields.
Since there are finitely many fields we are still able to obtain explicit formulas, without too much work.
The restricted Schur polynomials that span the su$(2|3)$ sector of the theory are labeled by 6 Young diagrams
and some multiplicity labels.
For operators dual to giant gravitons\cite{McGreevy:2000cw}, the Young diagram labels have a small number of long columns and for operators dual to dual giant gravitons\cite{Grisaru:2000zn,Hashimoto:2000zp}, the Young diagram labels have
a small number of long rows\cite{Balasubramanian:2001nh,CJR,David}.
These operators diagonalize the free field theory two point function to all orders in $1/N$ and they mix only weakly at
weak coupling.
In Section \ref{DonRSP} we derive an exact formula for the action of the one loop dilatation operator on restricted Schur
polynomials that span the su$(2|3)$ sector of the theory.
This is the first new result in this paper.
The novel ingredients involve the mixing of fermions, which was not considered in previous studies.
We find a rather simple way to express the complete result.
This result is exact in $1/N$.
By specializing to the operators dual to system of giant gravitons, in Section \ref{SimplD} we use simplifications of large $N$.
These simplifications suggest a new basis labeled by two Young diagrams and a graph, the so called Gauss graph
operators\cite{deMelloKoch:2012ck}.
The nodes of the graph correspond to the rows/columns of the Young diagram label.
There are also edges stretched between nodes in the graph and edges that have both end points on a given node.
We have derived a formula for the action of the dilatation operators in the Gauss graph basis of the su$(2|3)$ sector.
This is the second new result in this paper.
Matrix elements of the dilatation operator are given in terms of the number of edges between specific nodes on the graph.
Further, the dilatation operators preserves the number of edges stretched between nodes but can change the number of edges
with both endpoints attached to a given node.

This dilatation operator is rewritten in Section \ref{EmergentLattice} as a lattice model for particles.
The basic idea is simply to introduce oscillator creation and annihilation operators and then to rewrite the number of edges
in terms of these oscillators.
We demonstrate in Section \ref{EYM} that the resulting Hamiltonian is in detailed agreement with the Yang-Mills theory
expected as the world volume dynamics of the giant graviton branes.
Each Gauss graph operator becomes a state in a Fock space, with the graph giving an occupation number representation of the
states of the emergent world volume gauge theory.
This is the central result in this paper and it proves that the dilatation operator is mapped to the Hamiltonian of the dynamics
of the semi-classical physics of giant graviton branes.
Section \ref{discuss} contains some conclusions and a discussion of our results, which make a number of concrete suggestions.
For example, to explore the thermodynamics of the gravity theory dual to these large dimension operators, we argue that one
is considering the dynamics of a lattice model with all-to-all sites interactions and quenched disorder.
This looks a lot like the dynamics of the SYK model and our study may shed some light on the holographic relevance of models
with quenched disorder.

\section{Operators}\label{bases}

We use two bases of operators in this study.
A formula for matrix elements of the dilatation operator in the restricted Schur polynomial basis is the starting point for our study.
The formula we obtain is exact, meaning that it does not use any of the simplifications of large $N$.
Specializing to operators with bare dimension of order $N$, labeled by Young diagrams with order 1 long rows or columns,
naturally leads to a second basis for this class of operators, known as the Gauss graph operators.
Working in this basis allows us to exploit simplifications of large $N$.
Both bases are introduced and explained in this section.

\subsection{Restricted Schur Polynomials}

Restricted Schur polynomials\cite{Balasubramanian:2004nb,deMelloKoch:2007rqf,Bhattacharyya:2008rb}, have their genesis in
permutations groups and their representations, as well as in combinatorics of gauge invariant operators in multi-matrix models.
Although we will not use them in our study, note that closely related bases were introduced and studied
in \cite{Kimura:2007wy,Brown:2007xh,Brown:2008ij,Kimura:2008ac}.
Restricted Schur polynomials are labeled by a collection of Young diagrams and multiplicity labels, as we explain below.
They provide a basis for local gauge invariant operators of the theory, account for finite $N$ relations and diagonalize (to all
orders in $1/N$) the free field theory two point function\cite{Bhattacharyya:2008rb,Bhattacharyya:2008xy}.
When interactions are turned on, they only mix very weakly: at $L$-loops two operators will only mix if their labels differ by
moving at most $L$ boxes in any of the Young diagrams in the label\cite{DeComarmond:2010ie,Koch:2011hb}\footnote{See also \cite{deMelloKoch:2007nbd,Brown:2008rs}.}.
It is for these reasons that restricted Schur polynomials provide an attractive basis within which the operator mixing problem
can be formulated.

We truncate to the su$(2|3)$ sector of ${\cal N}=4$ super Yang-Mills theory.
Consequently, the Schur polynomials we study are constructed using three adjoint boson fields and two adjoint fermion fields.
Denote the bosonic fields by $\phi_i$ with $i=1,2,3$ and the fermionic fields by $\psi_a$ with $a=1,2$.
The complete set of gauge invariant observables is obtained by taking arbitrary products of these fields and then, to produce
a gauge invariant,  contracting row and column indices of the fields, in all possible ways.
Permuting row indices before tracing, we obtain all possible gauge invariant operators, with all possible trace structures.
We can label operators with the permutation that was performed on the row indices.
This labeling is redundant as a consequence of symmetries in the problem: swapping identical fields does not lead to
distinct observables.
Thus, there are two permutations groups that naturally enter the problem: the permutation group swapping row indices before
tracing and the permutation group swapping identical fields.
Consider operators constructed using $n_i$ of the $\phi_i$ fields and $m_a$ of the $\psi_a$ fields.
In what follows we will use the indices $A,\hat A$ running over the fields $A=\{\phi_1,\phi_2,\phi_3,\psi_1,\psi_2\}$ and
$\hat A=\{\phi_2,\phi_3,\psi_1,\psi_2\}$.
The permutation group swapping identical fields is given by the following product of symmetric groups
\bea
{\cal G}_{\rm symm}=S_{n_1}\times S_{n_2}\times S_{n_3}\times S_{m_1}\times S_{m_2}
\eea
The second permutation group that plays a role is the symmetric group $S_{n_T}$ which swaps row indices before tracing.
Here $n_T=n_1+n_2+n_3+m_1+m_2$ is the total number of fields appearing in the operator.
We must take the symmetry ${\cal G}_{\rm symm}$ into account to obtain a non-redundant labeling of the gauge
invariant operators.
This is done by recognizing that permutations labeling distinct observables belong to distinct restricted conjugacy
classes, as we now explain.
First, we will define the notion of a restricted conjugacy class: given a group $G$ and subgroup $H$, $g_1,g_2\in G$ are
in the same restricted conjugacy class if and only if $g_1=hg_2h^{-1}$ for some $h\in H$\cite{deMelloKoch:2007rqf}.
The operator corresponding to a permutation $\sigma$ is given by
\bea
&&{\rm Tr}\left(\sigma\, \psi^{\otimes m_1}_1 \psi^{\otimes m_2}_2
       \phi_3^{\otimes n_3} \phi_2^{\otimes n_2} \phi_1^{\otimes n_1}\right)
=\psi^{i_1}_{1\, i_{\sigma(1)}} \cdots \psi^{i_{m_1}}_{1\, i_{\sigma(m_1)}}
          \psi^{i_{m_1+1}}_{2\, i_{\sigma(m_1+1)}} \cdots \psi^{i_{m_1+m_2}}_{2\, i_{\sigma(m_1+m_2)}} \cr\cr
&&\qquad\qquad\qquad\times
          {\phi_3}^{i_{m_1+m_2+1}}_{i_{\sigma (m_1+m_2+1)}} \cdots
          {\phi_3}^{i_{m_1+m_2+n_3}}_{i_{\sigma (m_1+m_2+n_3)}}
          {\phi_2}^{i_{m_1+m_2+n_3+1}}_{i_{\sigma (m_1+m_2+n_3+1)}}\cdots
          {\phi_2}^{i_{m_1+m_2+n_3+n_2}}_{i_{\sigma (m_1+m_2+n_3+n_2)}}\cr\cr
&&\qquad\qquad\qquad\times
          {\phi_1}^{i_{m_1+m_2+n_3+n_2+1}}_{i_{\sigma (m_1+m_2+n_3+n_2+1)}}\cdots
          {\phi_1}^{i_{n_T}}_{i_{\sigma (n_T)}}
\eea
The $\psi_1$ fields are the first $m_1$ factors in the tensor product and the $\psi_2$ fields the next $m_2$ factors and so on.
We say that the $\psi_2$ fields, for example, occupy slots $m_1+1$ to $m_1+m_2$.
We will now argue that if we choose $G=S_{n_T}$ and $H={\cal G}_{\rm symm}$, then the difference between two
permutations in a given restricted conjugacy class is a permutation swapping identical fields so that they do indeed give
identical (possibly up to a sign for fermions) gauge invariant operators.
The result follows immediately after using the easily verified identity
\bea
{\rm Tr}(\rho^{-1}\sigma\rho A_1 A_2\cdots A_{n_T})
={\rm Tr}(\sigma A_{\rho (1)} A_{\rho (2)}\cdots A_{\rho (n_T)})
\eea
This does not quite remove the complete set of redundancies: observables that naively look independent can be linearly
dependent at finite $N$.
As an example, the Cayley-Hamilton Theorem tells us that any square matrix over a commutative ring satisfies its own
characteristic equation.
Taking a trace of this equation gives an identity between different multi trace structures, proving they are not linearly
independent.
To take these finite $N$ relations into account, perform a Fourier transform on the space of restricted conjugacy classes.
In the end, each field is in an irreducible representation of the permutation group permuting that species of field, and in an
irreducible representation of the permutation group permuting the entire collection of fields.
Since irreducible representations of permutation groups are labeled by Young diagrams there is one Young diagram label
for each species of field and one additional Young diagram for the complete collection of fields.
In addition, there are multiplicity labels.
These multiplicity labels are needed because upon restricting an irreducible representation of the group permuting the
complete set of fields in the operator, to the group that permutes only identical fields, many copies of a given irreducible
representation of the subgroup might arise.
Finite $N$ relations force polynomials labeled by a Young diagram with more than $N$ rows to vanish.
Thus, by keeping the restricted Schur polynomials, labeled by Young diagrams with at most $N$ rows, we obtain
a basis for the local gauge invariant operators.

For the su$(2|3)$ sector, the restricted Schur polynomials are given by \cite{Koch:2012sf}
\begin{eqnarray}
   \chi_{R,(\vec{r},\vec{s})\vec{\alpha}\vec{\beta}}(\phi_i,\psi_a)
   ={1\over n_1!n_2!n_3!m_1!m_2!}\sum_{\sigma\in S_{n_T}}
       \chi_{R,(\vec{r},\vec{s})\vec{\alpha}\vec{\beta}}(\sigma)
       {\rm Tr}\left(\sigma\, \psi^{\otimes m_1}_1 \psi^{\otimes m_2}_2
       \phi_3^{\otimes n_3} \phi_2^{\otimes n_2} \phi_1^{\otimes n_1}\right)
\label{su23restschur}
\end{eqnarray}
Each operator is indexed by a collection $R,(\vec{r},\vec{s})\vec{\alpha}\vec{\beta}$ of labels.
We know that swapping identical bosons is a symmetry, so we want a simultaneous swap of row and column indices of
bosons in the operator to leave the operator invariant.
The only way to get the invariant is to place row and column indices into the same representation $r$ and then project
to the (unique) invariant in $r\times r$.
Thus, the row and column indices of each bosonic field are in a definite representation.
The indices of the $\phi_i$ fields are in representation\footnote{The notation $r\vdash n$ means that $r$ is a partition of $n$.
Every Young diagram can be understood as a partition of an integer with the parts recording how many boxes
there are in each row of the diagram.
Consequently we write $r\vdash n$ to state that $r$ is a Young diagram with $n$ boxes.} $r_i\vdash n_i$.
For the fermionic fields we need to place the fermions into a totally antisymmetric representation, and this is achieved by
placing the row indices into some representation $s$ and the column indices into the conjugate representation $s^T$ and then
projecting to the (unique) antisymmetric representation appearing in $s\times s^T$ as explained in \cite{Koch:2012sf}.
The conjugate representation is obtained by flipping the Young diagram so that row and column lengths are exchanged.
We place the row indices of the $\psi_a$'s into representation $s_a\vdash m_1$ and the column indices into $s_a^T$.
The complete set of fields are in representation $R\vdash n_T$.
To refer to collections of Young diagrams we will use the notation $\vec r$ and $\vec s$, etc.
The collection $(\vec r,\vec s)$ specifies a representation of ${\cal G}_{\rm symm}$ which is a subgroup of $S_{n_T}$.
The representation of the subgroup is a subspace of the carrier space of representation $R$.
At this point we are forced to introduce multiplicities because the representation of the subgroup may appear more
than once.
We imagine embedding the subspace by removing $m_1$ boxes\footnote{Each box corresponds to a field. Thus,
each box corresponds to a row index and a column index and the collection of row and column indices must each be put into
an irreducible representation. This is why in the discussion that follows we assemble the boxes into two representations, each
with their own multiplicity label.} from $R$, and assembling them into representation $s_1$ and $s_1^T$.
There may be more than one way to do this, so that there may be more than one copy of these spaces.
Distinguish the different copies using the labels $\alpha_3$ (for $s_1$) and $\beta_3$ (for $s_1^T$).
Next $m_2$ boxes are removed and assembled into $s_2$ and $s_2^T$, with multiplicities $\alpha_4$ and $\beta_4$.
The next $n_3$ boxes are removed and assembled into $r_3$ with multiplicities $\alpha_2$ and $\beta_2$ and finally,
$n_2$ boxes are removed and assembled into $r_2$ with multiplicities $\alpha_1$ and $\beta_1$.
The last $n_1$ boxes remaining in $R$ are identified with $r_1$ so that $r_1$ is multiplicity free.

$\chi_{R,(\vec{r},\vec{s})\vec{\alpha}\vec{\beta}}(\sigma)$ is a restricted character \cite{deMelloKoch:2007nbd},
obtained by summing the row index of $\Gamma^R(\sigma)$ over the subspace $(\vec{r},\vec{s})\vec{\alpha}$
and the column index over the subspace $(\vec{r},\vec{s})\vec{\beta}$ which both arise upon restricting irreducible
representation $R$ of $S_{n_T}$ to its $S_{n_1}\times S_{n_2}\times S_{n_3}\times S_{m_1}\times S_{m_2}$ subgroup,
as we have just explained in detail.
The reader can consult \cite{Bhattacharyya:2008rb} for further details and results.
Here we simply note that the restricted characters are a complete set of functions on the restricted conjugacy classes, so
that the formula (\ref{su23restschur}) can be understood as a Fourier transform from the space of restricted
conjugacy classes, to the space of Young diagrams and multiplicity labels.
This interpretation relies on basic ideas first introduced in the pioneering paper \cite{Brown:2007xh}.
Even at finite $N$, restricted Schur polynomials are linearly independent \cite{Bhattacharyya:2008rb,Bhattacharyya:2008xy}
and diagonalize the free field theory two point function \cite{Bhattacharyya:2008rb}.
A straight forward computation now shows that
\bea
\langle \chi_{R,(\vec{r},\vec{s})\vec{\alpha}\vec{\beta}}(\phi_i,\psi_a)
        \chi^\dagger_{T,(\vec{t},\vec{u})\vec{\gamma}\vec{\delta}}(\phi_i,\psi_a)\rangle
=\delta_{R T}\delta_{\vec r\vec t}\delta_{\vec s\vec u}
\delta_{\vec\alpha\vec\gamma}\delta_{\vec\beta\vec\delta}
{f_R {\rm hooks}_{R}\over \prod_m {\rm hooks}_{r_m}\prod_n {\rm hooks}_{s_n}}
\eea
where we are using the obvious notation
\bea
\delta_{\vec r\vec t}=\prod_{i=1}^3\delta_{r_i t_i}\qquad
\delta_{\vec s\vec u}=\prod_{a=1}^2 \delta_{s_a u_a}\qquad
\delta_{\vec\alpha\vec\gamma}\delta_{\vec\beta\vec\delta}
=\prod_{k=1}^4 \delta_{\alpha_k\gamma_k}\delta_{\beta_k \delta_k}
\eea
Simple counting arguments prove that the number of restricted Schur polynomials matches the number of gauge
invariant operators that can be defined \cite{Koch:2012sf}.

In working with the restricted character it is useful to write
\bea
\chi_{R,(\vec{r},\vec{s})\vec{\alpha}\vec{\beta}}(\sigma)= {\rm Tr}_R
\left(P_{R,(\vec{r},\vec{s})\vec{\alpha}\vec{\beta}}\Gamma^{(R)}(\sigma)\right)
\eea
The trace is over the carrier space of irreducible representation $R$.
The operator $P_{R,(\vec{r},\vec{s})\vec{\alpha}\vec{\beta}}$ is an intertwining map, which sends the row
indices of $\Gamma^{(R)}(\sigma)$ to the $\vec\beta$ copy of $(\vec{r},\vec{s})$ and the column indices to the
$\vec\alpha$ copy of $(\vec{r},\vec{s})$ in the above trace.
The intertwining maps obey
\bea
&&{\rm sgn}(\rho_1){\rm sgn}(\rho_2)P_{R,(\vec{r},\vec{s})\vec{\alpha}\vec{\beta}}
\Gamma_{r_1}(\sigma_1)\otimes \Gamma_{r_2}(\sigma_2)\otimes
\Gamma_{r_3}(\sigma_3)\otimes \Gamma_{s_1^T}(\rho_1)\otimes \Gamma_{s_2^T}(\rho_2)\nonumber\\[1mm]
&&\qquad =\,\,\Gamma_{r_1}(\sigma_1)\otimes\Gamma_{r_2}(\sigma_2)\otimes
\Gamma_{r_3}(\sigma_3)\otimes\Gamma_{s_1}(\rho_1)\otimes\Gamma_{s_2}(\rho_2)
P_{R,(\vec{r},\vec{s})\vec{\alpha}\vec{\beta}}\label{InTMP}
\eea
as well as
\bea
P_{R_1,(\vec{r}_1,\vec{s}_1)\vec{\alpha}_1\vec{\beta}_1}
P_{R_2,(\vec{r}_2,\vec{s}_2)\vec{\alpha}_2\vec{\beta}_2}^\dagger
=\delta_{R_1 R_2}\delta_{\vec{r}_1\vec{r}_2}\delta_{\vec{s}_1\vec{s}_2}\delta_{\vec{\alpha}_2\vec{\beta}_1}
\bar P_{R_1,(\vec{r}_1,\vec{s}_1)\vec{\alpha}_1\vec{\beta}_2}\label{ProfPrrod}
\eea
and they can be written as a tensor product as follows
\bea
\bar P_{R,(\vec{r},\vec{s})\vec{\alpha}\vec{\beta}}&=&p_{r_1}\otimes p_{r_2\alpha_1\beta_1}\otimes
p_{r_3\alpha_2\beta_2}\otimes p_{s_1\alpha_3\beta_3} \otimes p_{s_2\alpha_4\beta_4}
\label{projtensorprod}
\eea

Finally, we find it convenient to work with restricted Schur polynomials normalized to have a unit two point function.
The normalized operators are defined by
\bea
  \chi_{R,(\vec{r},\vec{s})\vec{\alpha}\vec{\beta}}(\phi_i,\psi_a)
=\sqrt{f_R \, {\rm hooks}_R\over \prod_m {\rm hooks}_{r_m}\prod_n {\rm hooks}_{s_n}}
O_{R,(\vec{r},\vec{s})\vec{\alpha}\vec{\beta}}(\phi_i,\psi_a)\cr
\label{NrmlzdRSP}
\eea

\subsection{Gauss Graph Basis}\label{GGBasis}

We now specialize to operators which have a definite semi-classical limit in the holographically dual
theory\footnote{Specializing to classes of operators is always necessary.
Even in the planar limit one is forced to restrict attention to operators of dimension $\Delta$ with $\Delta^2/N\ll 1$.}.
Doing so will allow us to exploit the simplifications of large $N$ and, for this class of operators, a diagonalization of the
one loop dilatation operator.
The operators we consider have a dimension $\Delta\sim N$ so that the Young diagram $R$ labeling  the operator has
$\sim N$ boxes.
In addition $R$ has a fixed $\sim 1$ number of rows or columns.
Operators with $p$ long columns are dual to a system of $p$ giant gravitons and operators with $p$ long rows are dual
to $p$ dual giant gravitons\footnote{Branes connected by an open string described using a spin chain have been considered in
\cite{Berenstein:2013md,Berenstein:2013eya,Berenstein:2014isa,Berenstein:2014zxa,Koch:2015pga}.}.
These operators mix with each other, but not with operators labeled by Young diagrams that have a different number
of rows or columns.
In what follows, for simplicity we will discuss the case of long rows.
There is an identical discussion for long columns.
We consider operators constructed using mainly $\phi_1$ fields, so that $n_1\sim N$.
In addition, there are some bosonic $\phi_2,\phi_3$ excitations, as well fermionic $\psi_1,\psi_2$ excitations.
The number of excitations is limited by fixing $n_2\sim n_3\sim m_1\sim m_2\sim \sqrt N$.

A key observation motivating the Gauss graph basis concerns the shape of the $R$ Young diagram of the generic operator:
almost all operators in this class have unequal row lengths.
The difference in the length of any two distinct rows in $R$ is generically of size $aN$, where $a$ is a number
of order 1, possibly with $a\ll 1$.
The one loop dilatation operator moves a single box at a time so that of the order of $N$ applications are required to produce
operators with equal row lengths.
Thus, at weak coupling, if we start with sufficiently unequal lengths, we always have unequal lengths.
The conclusion is that, at large $N$ and weak coupling, corners on the right hand side of the Young diagram are well separated.
This limit was introduced and studied in \cite{Carlson:2011hy,Koch:2011hb} where it was called the displaced corners limit.
The action of the symmetric group on right most boxes simplifies in this limit: after neglecting order $1/N$ corrections,
permutations simply swap boxes they act on \cite{Koch:2011hb,deMelloKoch:2011qz}.
These are precisely the boxes that are to be removed and reassembled into irreducible representations of the subgroup which
is why this simplification has far reaching consequences.
The simplified action implies both new symmetries and new conservation laws.
Swapping row or column indices of fields of a given species, that belong to the same row, is a new symmetry.
The new conservation law manifests as the fact that operators only mix if they have the same number of excitation
fields of each species in a given row.
This new conservation law implies that we can refine the number of fields of a given species $N_{\hat A}$ to produce a $p$ dimensional vector
$\vec N_{\hat A}$, with each component recording how many fields are in a given row.
For example, the number of $\phi_2$ fields $n_2$ is refined to produce the vector $\vec n_2$.
The vector $\vec n_2$ labeling the dilatation operator is preserved so that an operator with vector $\vec n_2$ will not mix with
a second operator with $\vec n'_2$ if $\vec n_2\ne\vec n_2'$.
The group swapping $\phi_2$ fields in a given row, which is the enhanced symmetry of the displaced corners limit, is
\begin{equation}
   H_{\vec{n}_2}=S_{(n_2)_1}\times S_{(n_2)_2}\times \cdots\times S_{(n_2)_p}
\end{equation}
This symmetry acts independently on the row and column indices, so that the $\phi_2$ fields can be parametrized
by a permutation belonging to the double coset
\begin{equation}
   H_{\vec n_2}\setminus S_{n_2}/H_{\vec n_2}
\end{equation}
The number of values that the triple $(r_2,\alpha_1,\beta_1)$ takes equals the order of the double coset
$H_{\vec n_2}\setminus S_{n_2}/H_{\vec n_2}$, suggesting that instead of organizing the $\phi_2$ fields with the
$r_2,\alpha_1,\beta_1$ labels, we can organize them using the elements of the double coset \cite{deMelloKoch:2012ck}.
This is indeed the case, and the resulting basis is the Gauss graph basis.
The double cosets that are relevant for labeling our operators are given by
\bea
\phi_2&\leftrightarrow& \sigma_{\phi_2}\in H_{\vec{n}_2}\setminus S_{n_2}/H_{\vec{n}_2}\cr
\phi_3&\leftrightarrow& \sigma_{\phi_3}\in H_{\vec{n}_3}\setminus S_{n_3}/H_{\vec{n}_3}\cr
\psi_1&\leftrightarrow& \sigma_{\psi_1}\in H_{\vec{m}_1}\setminus S_{m_1}/H_{\vec{m}_1}\cr
\psi_2&\leftrightarrow& \sigma_{\psi_2}\in H_{\vec{m}_2}\setminus S_{m_2}/H_{\vec{m}_2}
\label{dcosets}
\eea
When we want to refer to a collection of permutations, one from each of the double cosets above, we will use the notation
$\vec\sigma=(\sigma_{\phi_2},\sigma_{\phi_3},\sigma_{\psi_1},\sigma_{\psi_2})$.

Gauss graph operators \cite{deMelloKoch:2012ck} are labeled by two Young diagrams (the $R$ and $r_1$ labels of the
restricted Schur polynomial) and a graph.
Nodes of the graph correspond to rows/columns of Young diagram $r_1$.
Each $\hat A$ field type ($\phi_2,\phi_3,\psi_1$ or $\psi_2$) corresponds to a species of edge in the graph and there
is an edge for each field.
The edges are directed and stretch between nodes.
An edge is allowed to leave and then return to the same node.
It is both convenient and possible to decompose the complete graph, to give a graph for each $\hat A$.
We can label the graphs using permutations, but this labeling is again redundant due to symmetries.
Swapping edges that terminate on a given node, or emanate from a given node is a symmetry.
This observation can be exploited to show that graphs are enumerated by elements of a double coset.
We refer the reader to \cite{deMelloKoch:2011uq} for the details.
So the appearance of double cosets in the displaced corners limit naturally leads to the graph labeling the operator.
The complete collection of graphs with $n$ edges and $p$ nodes, and with number of edges terminating at each node recorded
in $\vec n$ is described by a particular double coset.
The elements of the double cosets recorded in (\ref{dcosets}) correspond to the graphs we consider \cite{deMelloKoch:2011uq}.
We argue below that the number of edges give an occupation number description of the fields of the emergent gauge
theory defined on the world volume of the giant gravitons.
Consequently they must reflect constraints implied by the Gauss Law \cite{Balasubramanian:2004nb,Sadri:2003mx} which
manifests as the fact that only graphs with the same number of edges terminating on a node as number of edges emanating
from a node, for each species, are allowed.
This is the origin of the name Gauss graph \cite{deMelloKoch:2012ck}.
Fermi statistics forbids two or more parallel edges (i.e. edges with the same orientation and endpoints) of the same
fermion species\cite{deCarvalho:2020pdp}.
We refined $N_{\hat A}$ to produce a vector $\vec N_{\hat A}$.
To describe the graph we refine $\vec N_{\hat A}$ further to produce a matrix $(N_{\hat A})_{i\to j}$ whose elements
describe the number of edges running from node $i$ to node $j$.
We will abbreviate $(N_{\hat A})_{i\to i}$ as $(N_{\hat A})_{ii}$.
The total number of edges between nodes $i$ and $j$ is given by
$(N_{\hat A})_{ij}=(N_{\hat A})_{i\to j}+(N_{\hat A})_{j\to i}$.

The orthogonal transformation from the restricted Schur polynomial basis to the Gauss graph basis uses two types of
group theoretic coefficients.
The first set of coefficients
\bea
  C^{(r)}_{\mu_1\mu_2}(\tau)=|H_{\vec{n}}|\sqrt{d_{r}\over n!}\sum_{k,m=1}^{d_r}\Gamma^{(r)}(\tau)_{km}
   B^{r\to 1_{H_{\vec{n}}}}_{k\mu_1}B^{r\to 1_{H_{\vec{n}}}}_{m\mu_2}
\eea
are used to transform the labels of the $\phi_2,\phi_3$ fields.
In this formula $d_r$ is the dimension of irreducible representation $r\vdash n$ of $S_n$, $\Gamma^{(r)}(\tau)_{km}$ is
the matrix representing $\tau\in S_n$ in irreducible representation $r$ and
\bea
|H_{\vec n}|=n_1!n_2!\cdots n_p!
\eea
is the order of the group $H_{\vec n}$.
Finally, $B^{r\to 1_{H_{\vec{n}}}}_{k\mu_1}$ is a branching coefficient, described in more detail below.
The second set of group theoretic coefficients, distinguished by a tilde,
\bea
  \tilde C^{(s)}_{\mu_1\mu_2}(\tau)=|H_{\vec{m}}|\sqrt{d_s\over m!}
   \sum_{k,m=1}^{d_s}\left(\Gamma^{(s)}(\tau)\hat{O}\right)_{km}
   B^{s\to 1_{H_{\vec{m}}}}_{k\mu_1}B^{s^T\to 1^m_{H_{\vec m}}}_{m\mu_2}
\eea
are used to transform the $\psi_1,\psi_2$ labels.
We have introduced another set of branching coefficients, which are also discussed in more detail below, as well as
an operator $\hat O$, which maps from irreducible representation $s^T$ to $s$ and is normalized so that (here $1$
is the identity permutation)
\bea
\hat O^T\hat O=\Gamma^{(s^T)}(1)
\eea
In terms of these coefficients, the Gauss graph operators are
\bea
  O_{R,r_1}(\vec{\sigma})=\sum_{r_2\vdash n_2}\sum_{r_3\vdash n_3}\sum_{s_1\vdash m_1}\sum_{s_2\vdash m_2}\sum_{\vec{\mu},\vec{\nu}}
C^{(r_2)}_{\mu_1\nu_1}(\sigma_{\phi_2})
C^{(r_3)}_{\mu_2\nu_2}(\sigma_{\phi_3})
\tilde C^{(s_1)}_{\mu_3\nu_3}(\sigma_{\psi_1})
\tilde C^{(s_2)}_{\mu_4\nu_4}(\sigma_{\psi_2})
O_{R,(\vec{r},\vec{s})\vec{\mu}\vec{\nu}}\cr
\eea
In performing the basis change, the basic formulas that we need are properties of the branching coefficients
which we will now review.
The branching coefficients introduced above are defined by
\bea
  \sum_{\mu}B^{s\to 1_H}_{k\mu}B^{s\to 1_H}_{l\mu} = {1\over |H|}\sum_{\gamma\in H}\Gamma^{(s)}(\gamma)_{kl}
\eea
\bea
  \sum_{\mu}B^{s^T\to 1^m}_{k\mu}B^{s^T\to 1^m}_{l\mu} = {1\over |H|}\sum_{\gamma\in H}{\rm sgn}(\gamma)\Gamma^{(s^T)}(\gamma)_{kl}
\eea
The coefficients $B^{s\to 1_H}_{l\mu}$ resolve the multiplicities that arise when we restrict irreducible representation $s$
of $S_m$ to the identity representation $1_H$ of $H$ for which $\Gamma^{1_H}(\gamma)=1$ $\forall\gamma$.
The coefficients $B^{s\to 1^m}_{l\mu}$ resolve the multiplicities that arise when we restrict irreducible representation $s$
of $S_m$ to representation $1^m$ of $H$ for which $\Gamma^{1^m} (\gamma)={\rm sgn}(\gamma)$ $\forall\gamma$.
These branching coefficients are not independent: $B^{s\to 1_H}_{n\mu}O_{nl}=B^{s^T\to 1^m}_{l\mu}$.
This relation between the two implies that the transformation to Gauss graph basis is exactly the same for the fermions
and bosons \cite{deCarvalho:2020pdp}
\bea
  \tilde C^{(s_i)}_{\mu_1\mu_2}(\tau)=C^{(s_i)}_{\mu_1\mu_2}(\tau)
\eea

When evaluating matrix elements of the dilatation operator in the Gauss graph basis, it is convenient to work with operators
$\hat O_{R,r}(\vec\sigma)$ normalized to have a unit two point function.
They are related to the operators we have just defined as follows
\bea
O_{R,r}(\vec\sigma)=\sqrt{\prod_{\hat A=1}^4 \prod_{i,j=1}^p (N_{\hat A})_{i\to j}!}\,\,\, \hat O_{R,r}(\vec\sigma)
\eea

\section{Action of the dilatation operator on restricted Schur polynomials}\label{DonRSP}

The one loop dilatation operator in the su$(2|3)$ sector is given by \cite{Beisert:2003ys,Eden:2004ua}
\bea
  D=&-&{2g_{YM}^2\over (4\pi)^2}\left(\sum_{i>j=1}^3 \, {\rm Tr}\left(\left[\phi_i,\phi_j\right]\left[\partial_{\phi_i},\partial_{\phi_j}\right]\right)
     +\sum_{i=1}^3\sum_{a=1}^2 \, {\rm Tr}\left(\left[\phi_i,\psi_a\right]\left[\partial_{\phi_i},\partial_{\psi_a}\right]\right)\right.
\cr
&+&\,{\rm Tr}\left(\left\{\psi_1,\psi_2\right\}\left\{\partial_{\psi_1},\partial_{\psi_2}\right\}\right)
\Bigg)
\label{fullD}
\eea
It is useful to introduce the notation
\begin{eqnarray}
D\equiv -{2g_{YM}^2\over (4\pi)^2}\sum_{A>B=1}^5 D_{AB}
\end{eqnarray}
where $D_{AB}$ mixes fields of species $A$ and $B$.
To derive the action of $D$ on the restricted Schur polynomials, we need to evaluate the derivatives and then express the
result as a linear combination of restricted Schur polynomials.
The second step is always possible because the restricted Schur polynomials provide a basis.
In practice it is carried out using properties of restricted characters that imply \cite{Bhattacharyya:2008xy}
\bea
{\rm Tr}(\sigma \psi^{\otimes\, m_1}_1 \psi^{\otimes\, m_2}_2 \phi_1^{\otimes \, n_1}
\phi_2^{\otimes \, n_2} \phi_3^{\otimes \, n_3})=\sum_{R,(\vec r,\vec s)\vec\alpha\vec\beta}
{d_R n_1!n_2!n_3!m_1!m_2!\over d_{r_1}d_{r_2}d_{r_3}d_{s_1}d_{s_2}n_T!}
\chi_{R,(\vec r,\vec s)\vec\alpha\vec\beta} (\sigma^{-1})\chi_{R,(\vec r,\vec s)\vec\beta\vec\alpha} (\phi_i,\psi_a)\cr
\label{idnty}
\eea
Matrix elements arising from the mixing of two bosonic fields have been derived in \cite{DeComarmond:2010ie},
while matrix elements for the mixing of a boson and fermion field were derived in \cite{Koch:2012sf}.
Matrix elements relevant for the mixing of two fermionic fields have not been considered previously so we
will discuss their derivation in detail below.
To simplify the notation, introduce the following shorthand
\bea
&&1_{\psi_1}=1\qquad\qquad\qquad\qquad\qquad\,\,\,\,\,\,\, m_{\psi_1}=m_1\cr
&&1_{\psi_2}=m_1+1\qquad\qquad\qquad\qquad\,\,\,\, m_{\psi_2}=m_1+m_2\cr
&&1_{\phi_3}=m_1+m_2+1\qquad\qquad\qquad\,\, n_{\phi_3}=m_1+m_2+n_3\cr
&&1_{\phi_2}=m_1+m_2+n_3+1\qquad\qquad\, n_{\phi_2}=m_1+m_2+n_3+n_2\cr
&&1_{\phi_1}=m_1+m_2+n_3+n_2+1\qquad n_{\phi_1}=m_1+m_2+n_3+n_2+n_1=n_T
\eea
Due to the presence of fermionic fields we need to be careful about signs.
To evaluate the derivatives we need to compute
\bea
&&\{\psi_1,\psi_2\}^i_j\left({d\over d\psi_1{}^k_j}
{d\over d\psi_2{}{}^i_k}+{d\over d\psi_2{}^k_j}{d\over d\psi_1{}^i_k}\right)
   \sum_{\sigma\in S_{n_T}}{\rm Tr}_{(\vec r,\vec s)\vec{\mu}\vec{\nu}}(\Gamma^{(R)}(\sigma))
   \psi^{i_{1_{\psi_1}}}_{1\, i_{\sigma(1_{\psi_1})}}\cdots \psi^{i_{m_{\psi_1}}}_{1\, i_{\sigma(m_{\psi_1})}}\cr\cr
&&\qquad\times      \psi^{i_{1_{\psi_2}}}_{2\, i_{\sigma(1_{\psi_2})}}\cdots
                   \psi^{i_{m_{\psi_2}}}_{2\, i_{\sigma(m_{\psi_2})}}
    {\phi_3}^{i_{1_{\phi_3}}}_{i_{\sigma(1_{\phi_3})}}\cdots
    {\phi_3}^{i_{n_{\phi_3}}}_{i_{\sigma(n_{\phi_3})}}\cr
&&\qquad\times
    {\phi_2}^{i_{1_{\phi_2}}}_{i_{\sigma(1_{\phi_2})}}\cdots
    {\phi_2}^{i_{n_{\phi_2}}}_{i_{\sigma(n_{\phi_2})}}
    {\phi_1}^{i_{1_{\phi_1}}}_{i_{\sigma(1_{\phi_1})}}\cdots
    {\phi_1}^{i_{n_{\phi_1}}}_{i_{\sigma(n_{\phi_1})}}\cr\cr
&&=m_1m_2 \sum_{\sigma\in S_{n_T}}{\rm Tr}_{(\vec r,\vec s)\vec{\mu}\vec{\nu}}
(\Gamma^{(R)}([(1_{\psi_2},1),\sigma])) (-1)^{m_{1}}
\{\psi_1,\psi_2\}^{i_{1_{\psi_1}}}_{i_{\sigma(1_{\psi_1})}}
\psi^{i_{1+1_{\psi_1}}}_{1\, i_{\sigma(1+1_{\psi_1})}}\cdots
\psi^{i_{m_{\psi_1}}}_{1\, i_{\sigma(m_{\psi_1})}}\cr
 &&\qquad \times\delta^{i_{1_{\psi_2}}}_{i_{\sigma(1_{\psi_2})}}
        \psi^{i_{1_{\psi_2}+1}}_{2\, i_{\sigma(1_{\psi_2}+1)}}\cdots
    \psi^{i_{m_{\psi_2}}}_{2\, i_{\sigma(m_{\psi_2})}}
    {\phi_3}^{i_{1_{\phi_3}}}_{i_{\sigma(1_{\phi_3})}}\cdots
    {\phi_3}^{i_{n_{\phi_3}}}_{i_{\sigma(n_{\phi_3})}}\cr
&&\qquad\times
    {\phi_2}^{i_{1_{\phi_2}}}_{i_{\sigma(1_{\phi_2})}}\cdots
    {\phi_2}^{i_{n_{\phi_2}}}_{i_{\sigma(n_{\phi_2})}}
    {\phi_1}^{i_{1_{\phi_1}}}_{i_{\sigma(1_{\phi_1})}}\cdots
    {\phi_1}^{i_{n_{\phi_1}}}_{i_{\sigma(n_{\phi_1})}}
\nonumber
\eea
It will prove to be useful to have both indices of the Kronecker delta in the first slot, as this will allow us to express
the sum of $S_{n_T}$ as a sum over the subgroup $S_{n_T-1}$ and its cosets.
To achieve this, change summation variables from $\sigma$ to $\rho$ where $\sigma = (1,1_2)\rho (1,1_2)$ and
then relabel the summation variable back to the original name $\sigma$.
The result is
\bea
&&=m_1m_2\sum_{\sigma\in S_{n_T}}{\rm Tr}_{(\vec r,\vec s)\vec{\mu}\vec{\nu}}
(\Gamma^{(R)}([\sigma,(1,1_{\psi_2})]))\delta^{i_1}_{i_{\sigma(1)}}(-1)^{m_1}
\{\psi_1,\psi_2\}^{i_{1_{\psi_2}}}_{i_{\sigma(1_{\psi_2})}}\cr
&&\qquad\times
\psi^{i_{1+1_{\psi_1}}}_{1\, i_{\sigma(1+1_{\psi_1})}}\cdots \psi^{i_{m_{\psi_1}}}_{1\, i_{\sigma(m_{\psi_1})}}
\psi^{i_{1_{\psi_2}+1}}_{2\, i_{\sigma(1_{\psi_2}+1)}}\cdots
\psi^{i_{m_{\psi_2}}}_{2\, i_{\sigma(m_{\psi_2})}}
    {\phi_3}^{i_{1_{\phi_3}}}_{i_{\sigma(1_{\phi_3})}}\cdots
    {\phi_3}^{i_{n_{\phi_3}}}_{i_{\sigma(n_{\phi_3})}}\cr
&&\qquad\times
    {\phi_2}^{i_{1_{\phi_2}}}_{i_{\sigma(1_{\phi_2})}}\cdots
    {\phi_2}^{i_{n_{\phi_2}}}_{i_{\sigma(n_{\phi_2})}}
    {\phi_1}^{i_{1_{\phi_1}}}_{i_{\sigma(1_{\phi_1})}}\cdots
    {\phi_1}^{i_{n_{\phi_1}}}_{i_{\sigma(n_{\phi_1})}}\nonumber
\eea
Introduce the notation $\rho_i=\sigma (i,1)$ and rewrite the sum over $S_{n_T}$ as a sum over $S_{n_T-1}$ and its cosets.
The $S_{n_T-1}$ subgroup is obtained by restricting to permutations that leave 1 fixed, i.e. $\sigma(1)=1$.
The result is
\bea
&&=m_1m_2\sum_{\sigma\in S_{n_T-1}}\sum_{i=1}^{n_T}
{\rm Tr}_{(\vec r,\vec s)\vec{\mu}\vec{\nu}}
(\Gamma^{(R)}([\rho_i,(1,1_{\psi_2})]))\delta^{i_1}_{i_{\r_i(1)}}(-1)^{m_1}
\{\psi_1,\psi_2\}^{i_{1_{\psi_2}}}_{i_{\r_i(1_{\psi_2})}}
        \psi^{i_2}_{1\, i_{\r_i(2)}}\cdots \psi^{i_{m_1}}_{1\, i_{\r_i(m_1)}}\cr
&&\qquad\times \psi^{i_{1_{\psi_2}+1}}_{2\, i_{\r_i(1_{\psi_2}+1)}}\cdots
                   \psi^{i_{m_{\psi_2}}}_{2\, i_{\r_i(m_{\psi_2})}}
    {\phi_3}^{i_{1_{\phi_3}}}_{i_{\sigma(1_{\phi_3})}}\cdots
    {\phi_3}^{i_{n_{\phi_3}}}_{i_{\sigma(n_{\phi_3})}}
    {\phi_2}^{i_{1_{\phi_2}}}_{i_{\sigma(1_{\phi_2})}}\cdots
    {\phi_2}^{i_{n_{\phi_2}}}_{i_{\sigma(n_{\phi_2})}}
    {\phi_1}^{i_{1_{\phi_1}}}_{i_{\sigma(1_{\phi_1})}}\cdots
    {\phi_1}^{i_{n_{\phi_1}}}_{i_{\sigma(n_{\phi_1})}}\cr\cr
&&=m_1m_2 \sum_{\sigma\in S_{n_T-1}}{\rm Tr}_{(\vec r,\vec s)\vec{\mu}\vec{\nu}}
\Big(\Gamma^{(R)}([\s\Big\{N+\sum_{i=1}^{n_T}(i,1)\Big\},(1,1_{\psi_2})])\Big)\cr
 &&\qquad\qquad\times (-1)^{m_1}{\rm Tr}(\sigma\cdot \psi_1^{\otimes m_1-1}\{\psi_1,\psi_2\}
\psi_2^{\otimes m_2-1}{\phi_3}^{\otimes n_3} {\phi_2}^{\otimes n_2}{\phi_1}^{\otimes n_1})\cr\cr
&&=m_1m_2\sum_{R'}c_{RR'}\sum_{\sigma\in S_{n_T-1}}
{\rm Tr}_{(\vec r,\vec s)\vec{\mu}\vec{\nu}}
\Big(\Big[\Gamma^{(R')}(\s),\Gamma^{(R)}\left((1,1_{\psi_2})\right)\Big]\Big)\cr
 &&\qquad\qquad\times (-1)^{m_1}{\rm Tr}(\sigma\cdot \psi_1^{\otimes m_1-1}\{\psi_1,\psi_2\}
\psi_2^{\otimes m_2-1}\phi_3^{\otimes n_3} \phi_2^{\otimes n_2}\phi_1^{\otimes n_1})
\nonumber
\eea
We are summing over the subgroup $S_{n_T-1}$ of the group $S_{n_T}$.
After restriction to the subgroup the irreducible representation $R$ of $S_{n_T}$ gives all representations $R'$ obtained
by dropping a single box from $R$, such that the result is still a valid Young diagram.
After restricting each $R'$ appears exactly once.
In moving from the second last to the last line above we use the fact that $\sum_{i=1}^{n_T}(i,1)$ is a Jucys-Murphy
element, and the eigenvalues of these elements acting on any state in $R'$ is the factor of the box dropped from $R$
to obtain $R'$.
We denote the factor of this box by $c_{RR'}$.
Recall that the factor of the box in row $i$ and column $j$ is $N-i+j$.
For a discussion with all the details, the reader should consult Appendix B of \cite{deMelloKoch:2007rqf}.
Now, we can write (recall that $\sigma(1)=1$)
\bea
\{\psi_1,\psi_2\}^{i_{1_{\psi_2}}}_{i_{\sigma(1_{\psi_2})}}=
\psi^{i_{1_{\psi_2}}}_{1\,i_{\sigma(1)}}\psi_2{}^{i_1}_{i_{\sigma(1_{\psi_2})}}
+\psi_2{}^{i_{1_{\psi_2}}}_{i_{\sigma(1)}}\psi^{i_1}_{1\, i_{\sigma(1_{\psi_2})}}
=\psi^{i_{1_{\psi_2}}}_{1\,i_{\sigma(1)}}\psi_2{}^{i_1}_{i_{\sigma(1_{\psi_2})}}
-\psi^{i_1}_{1\, i_{\sigma(1_{\psi_2})}}\psi_2{}^{i_{1_{\psi_2}}}_{i_{\sigma(1)}}
\eea
Consequently
\bea
&&(-1)^{m_1}{\rm Tr}(\sigma\cdot \psi_1^{\otimes m_1-1}\{\psi_1,\psi_2\}
\psi_2^{\otimes m_2-1}\phi_3^{\otimes n_3} \phi_2^{\otimes n_2}\phi_1^{\otimes n_1})\cr\cr
&&\qquad\qquad={\rm Tr}([(1,1_{\psi_2}),\sigma]\cdot \psi_1^{\otimes m_1}
\psi_2^{\otimes m_2} \phi_3^{\otimes n_3}\phi_2^{\otimes n_2}\phi_1^{\otimes n_1})
\eea
Thus, we now have
\bea
=m_1 m_2\sum_{R'}c_{RR'}\sum_{\sigma\in S_{n_T-1}}
{\rm Tr}_{(\vec r,\vec s)\vec{\mu}\vec{\nu}}
\Big(\Big[\Gamma^{(R')}(\s),\Gamma^{(R)}\left((1,1_{\psi_2})\right)\Big]\Big)\cr
{\rm Tr}(\big[(1,1_{\psi_2}),\sigma\big] \cdot \psi_1^{\otimes m_1}
\psi_2^{\otimes m_2} \psi_1 \phi_3^{\otimes n_3}\phi_2^{\otimes n_2}\phi_1^{\otimes n_1})
\nonumber
\eea
At this point use (\ref{idnty}) to obtain
\bea
 &&=\sum_{T,(\vec{t}\vec{u})\vec{\a}\vec{\b}}{m_1m_2d_T\over d_{t_1}d_{t_2}d_{t_3}d_{u_1}d_{u_2}n_T!}
\sum_{R'} c_{RR'}\cr
&&\qquad\qquad\sum_{\sigma\in S_{n_T-1}}
{\rm Tr}_{(\vec r,\vec s)\vec{\mu}\vec{\nu}}\Big(\Big[\Gamma^{(R')}(\s),\Gamma^{(R)}\left((1,1_2)\right)\Big]\Big)
{\rm Tr}_{(\vec{t}\vec{u})\vec{\a}\vec{\b}}\Big(\big[\sigma,(1,1_2)\big]\Big)
\chi_{T,(\vec t\vec u)\vec{\b}\vec{\a}}(\phi_i,\psi_a)\nonumber
\eea
The final step entails using Schur's orthogonality relations to perform the sum over the subgroup.
In the basis of normalized restricted Schur polynomials, the result is
\bea
D_{\psi_1\psi_2}O_{R,(\vec{r}\vec s)\vec\mu\vec\nu}=
\sum_{R'}\sum_{T,(\vec{t}\vec{u})\vec{\a}\vec{\b}}\sqrt{c_{RR'}c_{TT'}}
\sqrt{{\rm hooks}_{\vec r}{\rm hooks}_{\vec s}{\rm hooks}_{\vec t}{\rm hooks}_{\vec u}
\over {\rm hooks}_T{\rm hooks}_R}
{m_1m_2\sqrt{{\rm hooks}_{R'}{\rm hooks}_{T'}}\over n_1!n_2!n_3!m_1!m_2!}
\cr
\times{\rm Tr}_{R\oplus T}([P_{R,(\vec r\vec s)\vec{\mu}\vec{\nu}},\Gamma^{(R)}\left((1,1_2)\right)]I_{R'T'}
[P_{T,(\vec t\vec u)\vec{\a}\vec{\b}},\Gamma^{(T)}\left((1,1_2)\right)]I_{T'R'})
O_{T,(\vec t\vec u)\vec{\b}\vec{\a}}(\phi_i,\psi_a)\cr
\label{D12}
\eea
where
\bea
D_{\psi_1\psi_2}=\{\psi_1,\psi_2\}^i_j\left({d\over d\psi_1{}^k_j}{d\over d\psi_2{}^i_k}
+ {d\over d\psi_2{}^k_j}{d\over d\psi_1{}^i_k}\right)
\eea
In writing (\ref{D12}) we have introduce the intertwining map $I_{R'T'}$ which maps from the carrier space of $T'$ to the
carrier space of $R'$. This map vanishes if $R'$ and $T'$ do not have the same shape, which implies that the above matrix
element is non-zero if and only if $R$ and $T$ differ at most, by one box.
This map arises from the application of Schur's orthogonality relation, when performing the sum over the
$S_{n_T-1}$ subgroup.

Using the same procedure, all terms appearing in the dilatation operator can be evaluated.
In terms of the index $A=\{\phi_1,\phi_2,\phi_3,\psi_1,\psi_2\}$ introduced above, we write
$1_A=\{1_{\phi_1},1_{\phi_2},1_{\phi_3},1_{\psi_1},1_{\psi_2}\}$ and
$N_A=\{n_1,n_2,n_3,m_1,m_2\}$.
The action of the dilatation operator in the restricted Schur polynomial basis can be summarized as follows
\bea
&&DO_{R,(\vec{r}\vec s)\vec\mu\vec\nu}(\psi_a,\phi_i)=
-{2g_{YM}^2\over (4\pi)^2}
\sum_{A>B=1}^5 \sum_{T,(\vec{t}\vec{u})\vec{\a}\vec{\b}}
({\cal M}_{AB})_{R,(\vec{r}\vec s)\vec\mu\vec\nu , T,(\vec{t}\vec{u})\vec{\a}\vec{\b}}\,\,
O_{T,(\vec t\vec u)\vec{\b}\vec{\a}}(\psi_a,\phi_i)\label{fullfinald}
\eea
\bea
({\cal M}_{AB})_{R,(\vec{r}\vec s)\vec\mu\vec\nu , T,(\vec{t}\vec{u})\vec{\a}\vec{\b}}
&=&\sum_{R'}\sqrt{c_{RR'}c_{TT'}}
\sqrt{{\rm hooks}_{\vec r}{\rm hooks}_{\vec s}{\rm hooks}_{\vec t}{\rm hooks}_{\vec u}
\over {\rm hooks}_T{\rm hooks}_R}
{N_A N_B\sqrt{{\rm hooks}_{R'}{\rm hooks}_{T'}}\over n_1! n_2! n_3! m_1! m_2!}
\cr
&&{\rm Tr}_{R\oplus T}\Big(
[\Gamma^{(R)}\left((1,1_A)\right)P_{R,(\vec r\vec s)\vec{\m}\vec{\n}}\Gamma^{(R)}\left((1,1_A)\right),
\Gamma^{(R)}\left((1,1_B)\right)]I_{R'T'}\cr\cr
&&\qquad\qquad\qquad\times
[\Gamma^{(T)}\left((1,1_A)\right)P_{T,(\vec t\vec u)\vec{\a}\vec{\b}}\Gamma^{(T)}\left((1,1_A)\right),
\Gamma^{(T)}\left((1,1_B)\right)]I_{T'R'}\Big)\cr\cr
&& \label{AB}
\eea
This result is exact in $1/N$.
The fomulas (\ref{fullfinald}) and (\ref{AB}) give the complete action of the dilatation operator in the
su$(2|3)$ sector, which is one of the new results of this paper.

\section{Dilatation Operator on Gauss graphs}\label{SimplD}

In this section we specialize the discussion and focus on operators dual to giant graviton branes.
This class of operators can be described using the Gauss graph basis.
The basic result we achieve is a rewriting of the matrix elements of the dilatation operator in the Gauss graph basis.
Concretely this entails computing the following Fourier transform
\bea
(M_{AB})_{R,r_1,\vec\sigma_1 , T,t_1 ,\vec\sigma_2}=
\sum_{\substack{r_2,r_3,\vec s,\vec\mu\vec\nu\\t_2,t_3,\vec u,\vec\alpha\vec\beta}}
C^{(r_2,r_3,\vec s)}_{\vec\mu\vec\nu}(\vec\sigma_1)
C^{(t_2,t_3,\vec u)}_{\vec\alpha\vec\beta}(\vec\sigma_2)
({\cal M}_{AB})_{R,(\vec{r}\vec s)\vec\mu\vec\nu , T,(\vec{t}\vec{u})\vec{\a}\vec{\b}}
\label{ggtform}
\eea
where
\bea
C^{(r_2,r_3,\vec s)}_{\vec\mu\vec\nu}(\vec\sigma)=
C^{(r_2)}_{\mu_1\nu_1}(\sigma_{\phi_2})
C^{(r_3)}_{\mu_2\nu_2}(\sigma_{\phi_3})
\tilde C^{(s_1)}_{\mu_3\nu_3}(\sigma_{\psi_1})
\tilde C^{(s_2)}_{\mu_4\nu_4}(\sigma_{\psi_2})
\eea

Our first task is to simplify the trace
\bea
{\cal T}_{AB}&=&{\rm Tr}_{R\oplus T}\Big(
[\Gamma^{(R)}\left((1,1_A)\right)P_{R,(\vec r\vec s)\vec{\m}\vec{\n}}
\Gamma^{(R)}\left((1,1_A)\right),\Gamma^{(R)}\left((1,1_B)\right)]I_{R'T'}\cr\cr
&&\qquad\qquad\times
[\Gamma^{(T)}\left((1,1_A)\right)P_{T,(\vec t\vec u)\vec{\a}\vec{\b}}\Gamma^{(T)}\left((1,1_A)\right),
\Gamma^{(T)}\left((1,1_B)\right)]I_{T'R'})\label{tocompute}
\eea
which appears in the expression for the term in the one loop dilatation operator that mixes $A$ and $B$ type excitations.
Up to this point we have worked at one loop, but to all orders in $1/N$.
We will for the first time start to use some of the simplifications of large $N$ by working in the displaced corners approximation.

We start by introducing a vector space as explained in \cite{Koch:2011hb}.
Each box associated to an excitation becomes a $p$-dimensional vector in a space $V_p$.
Excitation boxes belonging to the $i$th row of $R$ are represented by vectors that have all entries equal to zero except for
the $i$th entry which is 1.
In this way the collection of impurities become a vector in $V_p^{\otimes n_2+n_3+m_1+m_2}$.
To explain the utility of this vector space, recall that each Young diagram $R$ can be labeled to produce a set of
Young-Yamanouchi (YY) symbols.
Each YY symbol corresponds to a state in the carrier space of irreducible representation $R$.
Translating each YY symbol into a vector in $V_p^{\otimes n_2+n_3+m_1+m_2}$, the action of the symmetric group
on $R$ becomes a simple action of permuting vectors in $V_p^{\otimes n_2+n_3+m_1+m_2}$.
For a detailed account of this mathematical framework the reader should consult \cite{deMelloKoch:2011qz}.

The calculations of this section make extensive use of (\ref{projtensorprod}) which writes the intertwining map used to
construct the restricted Schur polynomial as a tensor product with a factor for each species of field.
The factor associated to the $\phi_1$ field is a projection operator.
The factors associated to excitation fields are themselves intertwining maps.

Imagine that $R'$ is obtained from $R$ by dropping a box in row $i$ and $T'$ from $T$ by dropping a box from row $j$.
The corresponding intertwining maps are
\bea
   I_{R'T'}=E^{(1)}_{ij},\qquad I_{T'R'}=E^{(1)}_{ji}
\eea
Here the $E_{ij}$ are the usual basis for $GL(N)$, i.e. $(E_{ij})_{ab}=\delta_{ia}\delta_{jb}$.
The superscript tells us which factor in the tensor product $E_{ij}$ acts on.
To evaluate the traces, write the permutations appearing in the above trace (\ref{tocompute}) in terms of the $E_{ij}$'s
\bea
(1_A,1_B)=\sum_{i,j=1}^p E^{(1_A)}_{ij}E^{(1_B)}_{ji}
\eea
and simplify the product of the $E$'s using the usual algebra
\bea
E_{ij}E_{kl}=\delta_{jk}E_{il}
\eea
By the usual rules for the tensor product, $E$s only multiply with each other if they are in the same slot.
Since the trace ${\cal T}$ is a product of two commutators, expanding gives four terms.
After expanding we have
\bea
&&{\cal T}_{AB}=
 \left( {\rm Tr}_{R\oplus T}\big(
(1,1_A)P_{R,(\vec r\vec s)\vec{\m}\vec{\n}}(1,1_A)(1,1_B)I_{R'T'}
(1,1_A)P_{T,(\vec t\vec u)\vec{\a}\vec{\b}}(1,1_A)(1,1_B)I_{T'R'}\big)\right.\cr
&&         -{\rm Tr}_{R\oplus T}\big(
(1,1_B)(1,1_A)P_{R,(\vec r\vec s)\vec{\m}\vec{\n}}(1,1_A)I_{R'T'}
(1,1_A)P_{T,(\vec t\vec u)\vec{\a}\vec{\b}}(1,1_A)(1,1_B)I_{T'R'}\big)\cr
&&         -{\rm Tr}_{R\oplus T}\big(
(1,1_A)P_{R,(\vec r\vec s)\vec{\m}\vec{\n}}(1,1_A)(1,1_B)I_{R'T'}
(1,1_B)(1,1_A)P_{T,(\vec t\vec u)\vec{\a}\vec{\b}}(1,1_A)I_{T'R'}\big)\cr
&&\left.
           +{\rm Tr}_{R\oplus T}\big(
(1,1_B)(1,1_A)P_{R,(\vec r\vec s)\vec{\m}\vec{\n}}(1,1_A)I_{R'T'}
(1,1_B)(1,1_A)P_{T,(\vec t\vec u)\vec{\a}\vec{\b}}(1,1_A)I_{T'R'}\big)\right)
\eea
Following the procedure described above we find
\bea
(1,1_A)(1,1_B)I_{R'T'}(1,1_A)&=&(1,1_A)(1,1_B)E^{(1)}_{ij}(1,1_A)=(1_A,1_B)E^{(1_A)}_{ij}\cr
(1,1_A)I_{R'T'}(1,1_A)&=&(1,1_A)E^{(1)}_{ij}(1,1_A)=E^{(1_A)}_{ij}\cr
(1,1_A)(1,1_B)I_{T'R'}(1,1_B)(1,1_A)&=&(1,1_A)(1,1_B)E^{(1)}_{ji}(1,1_B)(1,1_A)=E^{(1_B)}_{ji}
\eea
Thus, the trace now simplifies to
\bea
&&{\cal T}_{AB}=
\left( {\rm Tr}_{R\oplus T}\big(
P_{R,(\vec r\vec s)\vec{\m}\vec{\n}}E^{(1_B)}_{ia}E^{(1_A)}_{aj}
P_{T,(\vec t\vec u)\vec{\a}\vec{\b}}E^{(1_B)}_{jc}E^{(1_A)}_{ci}\big)\right.\cr
&&         -{\rm Tr}_{R\oplus T}\big(
P_{R,(\vec r\vec s)\vec{\m}\vec{\n}}E^{(1_A)}_{ij}P_{T,(\vec t\vec u)\vec{\a}\vec{\b}}E^{(1_B)}_{ji}\big)
 -{\rm Tr}_{R\oplus T}\big(
P_{R,(\vec r\vec s)\vec{\m}\vec{\n}}E^{(1_B)}_{ij}P_{T,(\vec t\vec u)\vec{\a}\vec{\b}}E^{(1_A)}_{ji}\big)\cr
&&\left.
           +{\rm Tr}_{R\oplus T}\big(
P_{R,(\vec r\vec s)\vec{\m}\vec{\n}}E^{(1_A)}_{ib}E^{(1_B)}_{bj}
P_{T,(\vec t\vec u)\vec{\a}\vec{\b}}E^{(1_A)}_{jd}E^{(1_B)}_{di}\big)\right)
\eea
With this simplified expression in hand we can return to evaluating the sums in (\ref{ggtform}).

The terms involving mixing of excitations with $\phi_1$ are significantly simpler due to the fact that the projector
is simpler. These terms have already been evaluated for bosons in \cite{deMelloKoch:2012ck} and for fermions in
\cite{Koch:2012sf}.
The result is (we remind the reader that the integers $(n_2)_{ij}$, $(n_3)_{ij}$, $\cdots$ were defined in the paragraph
after the paragraph containing (\ref{dcosets}))
\bea
\label{ggaction}
D_{\phi_1\phi_2} O_{R,r_1}(\vec\sigma)
= \sum_{i>j=1}^p(n_2)_{ij}\Delta_{ij}O_{R,r_1}(\vec\sigma)\cr
D_{\phi_1\phi_3} O_{R,r_1}(\vec\sigma)
= \sum_{i>j=1}^p(n_3)_{ij}\Delta_{ij}O_{R,r_1}(\vec\sigma)\cr
D_{\phi_1\psi_1} O_{R,r_1}(\vec\sigma)
= \sum_{i>j=1}^p(m_1)_{ij}\Delta_{ij}O_{R,r_1}(\vec\sigma)\cr
D_{\phi_1\psi_2} O_{R,r_1}(\vec\sigma)
= \sum_{i>j=1}^p(m_2)_{ij}\Delta_{ij}O_{R,r_1}(\vec\sigma)
\eea
The operator $\Delta_{ij}$ is a sum of three terms
\bea
\Delta_{ij}=\Delta_{ij}^{+}+\Delta_{ij}^{0}+\Delta_{ij}^{-}
\eea
$\Delta_{ij}$ acts only on the $R,r_1$ labels of the Gauss graph operator.
Denote the row lengths of Young diagram $r$ by $l_r$.
Young diagram $r_{ij}^+$ is obtained by removing a box from row $j$ and adding it to row $i$
and $r_{ij}^-$ is obtained by removing a box from row $i$ and adding it to row $j$.
With this notation the action of the terms appearing in $\Delta_{ij}$ are
\begin{eqnarray}
  \Delta_{ij}^{0}O_{R,r}(\vec\sigma) &=& -(2N+l_{r_i}+l_{r_j})O_{R,r}(\vec\sigma)\cr
  \Delta_{ij}^{\pm}O_{R,r}(\vec\sigma)&=&\sqrt{(N+l_{r_i})(N+l_{r_j})}O_{R^\pm_{ij},r^\pm_{ij}}(\vec\sigma)
\end{eqnarray}

Now consider the contributions $D_{\hat A\hat B}$ to the dilatation operator that describe the mixing of the excitation fields.
These terms share the same structure, so we can carry the discussion out in generality.
Using the results of \cite{Koch:2013yaa}, we obtain the following expression
\bea
D_{\hat A\hat B} \hat O_{R,r_1}(\vec\sigma_1)= (M_{\hat A\hat B})_{R,r_1,\vec\sigma_1\,\,T,t_1,\vec\sigma_2}
\hat O_{T,t_1}(\vec\sigma_2)
\label{exdil}
\eea
where
\bea
\label{MinDA}
&&(M_{\hat A\hat B})_{R,r_1,\vec\sigma_1\,\,T,t_1,\vec\sigma_2}=
{1\over\sqrt{|O_{R,r_1}(\vec\sigma_1)|^2 |O_{T,t_1}(\vec\sigma_2)|^2}}
\prod_{\hat C\ne\hat A,\hat B}\delta_{(\sigma_{\hat C})_1(\sigma_{\hat C})_2}
\times\cr\cr
&&\sum_{R'}{\delta_{R'_iT'_k}\delta_{r_1t_1}\over (N_{\hat A}-1)!(N_{\hat B}-1)!}
\sqrt{c_{RR'}c_{TT'}\over l_{R_i}l_{T_k}}
\sum_{\psi_1\in S_{N_{\hat A}}\times S_{N_{\hat B}}}\sum_{\psi_2\in S_{N_{\hat A'}}\times S_{N_{\hat B'}}}\cr\cr
&&\Big[
\langle\vec{N}_{\hat A}',\vec{N}_{\hat B}'|\sigma_2\psi_2^{-1}E^{(1_{\hat A})}_{ki}\psi_1
|\vec{N}_{\hat A},\vec{N}_{\hat B}\rangle
\langle \vec{N}_{\hat A},\vec{N}_{\hat B}|\sigma_1^{-1}\psi_1^{-1}E^{(1_{\hat B})}_{ik}\psi_2
|\vec{N}_{\hat A}',\vec{N}_{\hat B}'\rangle\cr\cr
&&-\langle\vec{N}_{\hat A}',\vec{N}_{\hat B}'|\sigma_2\psi_2^{-1}E^{(1_{\hat A})}_{ci}E^{(1_{\hat B})}_{kc}\psi_1
|\vec{N}_{\hat A},\vec{N}_{\hat B}\rangle
\langle \vec{N}_{\hat A},\vec{N}_{\hat B}|\sigma_1^{-1}\psi_1^{-1}E^{(1_{\hat A})}_{ak}E^{(1_{\hat B})}_{ia}\psi_2
|\vec{N}_{\hat A}',\vec{N}_{\hat B}'\rangle\cr\cr
&&-\langle\vec{N}_{\hat A}',\vec{N}_{\hat B}'|\sigma_2\psi_2^{-1}E^{(1_{\hat A})}_{kc}E^{(1_{\hat B})}_{ci}\psi_1
|\vec{N}_{\hat A},\vec{N}_{\hat B}\rangle
\langle\vec{N}_{\hat A},\vec{N}_{\hat B}|\sigma_1^{-1}\psi_1^{-1}E^{(1_{\hat A})}_{ia}E^{(1_{\hat B})}_{ak}\psi_2
|\vec{N}_{\hat A}',\vec{N}_{\hat B}'\rangle\cr\cr
&&+\langle\vec{N}_{\hat A}',\vec{N}_{\hat B}'|\sigma_2\psi_2^{-1}E^{(1_{\hat B})}_{ki}\psi_1
|\vec{N}_{\hat A},\vec{N}_{\hat B}\rangle
\langle\vec{N}_{\hat A},\vec{N}_{\hat B}|\sigma_1^{-1}\psi_1^{-1}E^{(1_{\hat A})}_{ik}\psi_2
|\vec{N}_{\hat A}',\vec{N}_{\hat B}'\rangle\Big]
\eea
Recall that the Gauss graph operators $\hat O_{R,r}(\vec\sigma_1)$ are normalized to have a unit two point function.
The delta function on the first line of the above expression vanishes if the graphs of the excitations that are not
mixing are not equal i.e. $\delta_{(\sigma_{\hat C})_1(\sigma_{\hat C})_2}=1$ as long as $(\sigma_{\hat C})_1$
and $(\sigma_{\hat C})_2$ correspond to the same double coset element.
The permutations $\sigma_1$ and $\sigma_2$ appearing in the above summand stand for the outer product of two
permutations.
Dropping the subscript for now, we can write $\sigma=\sigma_{\hat A}\circ\sigma_{\hat B}$ where ($\sigma$ should
not be confused with $\vec\sigma$)
\bea
\sigma_{\hat A}\in H_{\vec{N}_{\hat A}}\setminus S_{N_{\hat A}}/H_{\vec{N}_{\hat A}}\qquad\qquad
\sigma_{\hat B}\in H_{\vec{N}_{\hat B}}\setminus S_{N_{\hat B}}/H_{\vec{N}_{\hat B}}
\eea
We will also write this as
\bea
\sigma\in H\setminus S_{N_{\hat A}}\times S_{N_{\hat B}}/H\label{abcoset}
\eea
where $H=H_{\vec{N}_{\hat A}}\times H_{\vec{N}_{\hat B}}$.
We use $H_1$ for the symmetry group of $\sigma_1$ and $H_2$ for $\sigma_2$.
We use $N_{\hat A},N_{\hat B}$ for $\sigma_1$ and $N_{\hat A}',N_{\hat B}'$ for $\sigma_2$ and so on.
Some of this is simply for clarity: indeed, we always have $N_{\hat A}=N_{\hat A}'$ and $N_{\hat B}=N_{\hat B}'$,
but in general, $\vec{N}_{\hat A}\ne\vec{N}_{\hat A}'$ and $\vec{N}_{\hat B}\ne\vec{N}_{\hat B}'$.
We need to introduce the vectors $(\vec{v}_i)_a = \delta_{ia}$ which form a basis for $V_p$.
The vector $|\vec{N}_{\hat A},\vec{N}_{\hat B}\rangle$ is defined as follows
\bea
|\vec{N}_{\hat A},\vec{N}_{\hat B}\rangle = |\vec{N}_{\hat A}\rangle\otimes |\vec{N}_{\hat B}\rangle
\eea
where for any $p$ dimensional vector $\vec k$ we have
\bea
|\vec k\rangle &=& (\vec{v}_1)^{\otimes k_1}\otimes\cdots\otimes (\vec{v}_p)^{\otimes k_p}
\eea
With this notation in hand, we can now evaluate the sums over $\psi_1$ and $\psi_2$ in (\ref{MinDA}).
Consider the term
\bea
T_1=\sum_{\substack{\psi_1\in S_{N_{\hat A}}\times S_{N_{\hat B}}\\ \psi_2\in S_{N_{\hat A}'}\times S_{N_{\hat B}}'}}
\langle\vec{N}_{\hat A}',\vec{N}_{\hat B}'|\sigma_2\psi_2^{-1}E^{(1_{\hat A})}_{ki}\psi_1
|\vec{N}_{\hat A},\vec{N}_{\hat B}\rangle
\langle\vec{N}_{\hat A},\vec{N}_{\hat B}|\sigma_1^{-1}\psi_1^{-1}E^{(1_{\hat B})}_{ik}\psi_2
|\vec{N}_{\hat A}',\vec{N}_{\hat B}'\rangle
\nonumber
\eea
The dependence on the permutations $\sigma_1,\sigma_2$ can be simplified with the following
change of variables: replace $\psi_2$ with $\tilde\psi_2$ where
\bea
\tilde\psi_2=\psi_2\sigma_2^{-1}\qquad\Rightarrow\qquad \tilde\psi_2^{-1}=\sigma_2\psi_2^{-1}
\eea
After relabeling $\tilde\psi_2\to\psi_2$ and taking the transpose of the first factor which is a real number, we find
\bea
T_1=\sum_{\substack{\psi_1\in S_{N_{\hat A}}\times S_{N_{\hat B}}\\ \psi_2\in S_{N_{\hat A}'}\times S_{N_{\hat B}}'}}
\langle\vec{N}_{\hat A},\vec{N}_{\hat B}|\psi_1^{-1}E^{(1_{\hat A})}_{ik}\psi_2|\vec{N}_{\hat A}',\vec{N}_{\hat B}'\rangle
\langle\vec{N}_{\hat A},\vec{N}_{\hat B}|\sigma_1^{-1}\psi_1^{-1}E^{(1_{\hat B})}_{ik}\psi_2\sigma_2
|\vec{N}_{\hat A}',\vec{N}_{\hat B}'\rangle
\nonumber
\eea
If $i \ne k$, the first matrix element in the summand is only non-vanishing if $\vec{N}_{\hat A}\ne \vec{N}_{\hat A}'$ and
$\vec{N}_{\hat B}=\vec{N}_{\hat B}'$, while the second matrix element is only non-vanishing if
$\vec{N}_{\hat A}=\vec{N}_{\hat A}'$ and $\vec{N}_{\hat B} \ne \vec{N}_{\hat B}'$.
Thus, $T_1$ vanishes for $i\ne k$, indicated explicitly as follows
\bea
T_1=\delta_{ik}\sum_{\psi_1,\psi_2\in S_{N_{\hat A}}\times S_{N_{\hat B}}}
\langle\vec{N}_{\hat A},\vec{N}_{\hat B}|\psi_1^{-1}E^{(1_{\hat A})}_{ii}\psi_2|\vec{N}_{\hat A},\vec{N}_{\hat B}\rangle
\langle\vec{N}_{\hat A},\vec{N}_{\hat B}|\sigma_1^{-1}\psi_1^{-1}E^{(1_{\hat B})}_{ii}\psi_2\sigma_2
|\vec{N}_{\hat A},\vec{N}_{\hat B}\rangle
\nonumber
\eea
It makes sense to split the trace up as follows
\bea
T_1 &=& \delta_{ik}\sum_{\psi_1,\psi_2\in S_{N_{\hat A}}}
\langle\vec{N}_{\hat A}|\psi_1^{-1}E^{(1_{\hat A})}_{ii}\psi_2|\vec{N}_{\hat A}\rangle
\langle\vec{N}_{\hat A}|(\sigma_{\hat A})_1^{-1}\psi_1^{-1}\psi_2(\sigma_{\hat A})_2 |\vec{N}_{\hat A}\rangle\cr
&&\times\sum_{\psi_1,\psi_2\in S_{N_{\hat B}}}
\langle\vec{N}_{\hat B}|\psi_1^{-1}\psi_2|\vec{N}_{\hat B}\rangle
\langle\vec{N}_{\hat B}|(\sigma_{\hat B})_1^{-1}\psi_1^{-1}E^{(1_{\hat B})}_{ii}\psi_2(\sigma_{\hat B})_2
|\vec{N}_{\hat B}\rangle
\eea
Using the easily verified identity
\bea
\langle\vec{N}|\psi_1^{-1}\psi_2|\vec{N}\rangle=
\sum_{\rho\in H_{\vec N}}\delta(\rho\psi_1^{-1}\psi_2)
\eea
as well as $E^{(a)}_{ij}\psi_2=\psi_2E^{(\psi_2^{-1}(a))}_{ij}$ we find
\bea
T_1&=& \delta_{ik}\sum_{\psi_2\in S_{N_{\hat A}}}\sum_{\gamma\in H_{\vec{N}_{\hat A}}}
\langle\vec{N}_{\hat A}|(\sigma_{\hat A})_1\gamma^{-1}(\sigma_{\hat A})_2^{-1}E^{(\psi_2^{-1}(1_{\hat A}))}_{ii}
|\vec{N}_{\hat A}\rangle\cr
&&\times\sum_{\psi_2\in S_{N_{\hat B}}}\sum_{\rho\in H_{\vec{N}_{\hat B}}}
\langle\vec{N}_{\hat B}|(\sigma_{\hat B})_1^{-1}\rho^{-1}E^{(\psi_2^{-1}(1_{\hat B}))}_{ii}(\sigma_{\hat B})_2
|\vec{N}_{\hat B}\rangle\cr
&=& \delta_{ik}\sum_{\psi_2\in S_{N_{\hat A}}}\sum_{l\in S_i^{\hat A}}\delta(\psi_2^{-1}(1_{\hat A}),l)
\sum_{\gamma_1,\gamma_2\in H_{\vec{N}_{\hat A}}}
\delta ( (\sigma_{\hat A})_1\gamma^{-1}(\sigma_{\hat A})_2^{-1})\cr
&&\times\sum_{\psi_2\in S_{N_{\hat B}}}\sum_{l\in S_i^{\hat B}}\delta(\psi_2^{-1}(1_{\hat B}),l)
\sum_{\gamma_1,\gamma_2\in H_{\vec{N}_{\hat B}}}
\delta ((\sigma_{\hat B})_1\gamma^{-1}(\sigma_{\hat B})_2^{-1})\cr
&=& \delta_{ik}(N_{\hat A}-1)! (N^{\sigma_1}_{\hat A})_i \sum_{\gamma_1,\gamma_2\in H_{\vec{N}_{\hat A}}}
\delta ((\sigma_{\hat A})_1\gamma^{-1}(\sigma_{\hat A})_2^{-1})\cr
&&\qquad\qquad
\times (N_{\hat B}-1)! (N_{\hat B}^{\sigma_1})_i
\sum_{\gamma_1,\gamma_2\in H_{\vec{N}_{\hat B}}}
\delta ((\sigma_{\hat B})_1\gamma^{-1}(\sigma_{\hat B})_2^{-1})
\nonumber
\eea
and where $S_i^{\hat A}$ are the slots in $|\vec{N}_{\hat A}\rangle$ occupied by $\vec{v}_i$ and
$S_i^{\hat B}$ are the slots in $|\vec{N}_{\hat B}\rangle$ occupied by $\vec{v}_i$.
We can write this as
\bea
T_1=\delta_{ik}(N_{\hat A}-1)! (N_{\hat B}-1)! (\vec{N}_{\hat A})_i
(\vec{N}_{\hat B})_i\sum_{h_1,h_2\in H_2}
\delta(\sigma_1^{-1}h_1^{-1}\sigma_2h_2)
\eea
Here $(\vec{N}_{\hat A})_i$ and $(\vec{N}_{\hat B})_i$ count the number of edges ending on node $i$, or
equivalently the number of excitations of $\hat A/\hat B$ living in row $i$.
We could, for example, write
\bea
(\vec{N}_{\hat A})_i&=&\sum_{k\ne i}(N_{\hat A})_{k\to i}+(N_{\hat A})_{ii}\cr
&=&\sum_{k\ne i}(N_{\hat A})_{i\to k}+(N_{\hat A})_{ii}
\eea
where the second equality uses the constraints implied by the Gauss Law.

We also need to consider the term
\bea
T_4 &=&
\sum_{\substack{\psi_1\in S_{N_{\hat A}}\times S_{N_{\hat B}}\\ \psi_2\in S_{N_{\hat A}'}\times S_{N_{\hat B}'}}}
\langle\vec{N}_{\hat A}',\vec{N}_{\hat B}'|\sigma_2\psi_2^{-1}E^{(1_{\hat B})}_{ki}\psi_1
|\vec{N}_{\hat A},\vec{N}_{\hat B}\rangle
\langle \vec{N}_{\hat A},\vec{N}_{\hat B}|\sigma_1^{-1}\psi_1^{-1}E^{(1_{\hat A})}_{ik}\psi_2
|\vec{N}_{\hat A}',\vec{N}_{\hat B}'\rangle\cr
&=&\sum_{\substack{\psi_1\in S_{N_{\hat A}}\times S_{N_{\hat B}}\\ \psi_2\in S_{N_{\hat A}'}\times S_{N_{\hat B}'}}}
\langle \vec{N}_{\hat A},\vec{N}_{\hat B}|\sigma_1^{-1}\psi_1^{-1}E^{(1_{\hat A})}_{ik}\psi_2
|\vec{N}_{\hat A}',\vec{N}_{\hat B}'\rangle
\langle\vec{N}_{\hat A},\vec{N}_{\hat B}|\psi_1^{-1}E^{(1_{\hat B})}_{ik}\psi_2\sigma_2^{-1}
|\vec{N}_{\hat A}',\vec{N}_{\hat B}'\rangle\nonumber
\eea
Changing variables $\psi_1^{-1}\to\sigma_1^{-1}\psi_1^{-1}$ shows that $T_4 = T_1$ and hence
\bea
T_1 + T_4 = 2\delta_{ik}(N_{\hat A}-1)! (N_{\hat B}-1)! (\vec{N}_{\hat A})_i
(\vec{N}_{\hat B})_i\sum_{h_1,h_2\in H_2}\delta(\sigma_1^{-1}h_1^{-1}\sigma_2h_2)
\eea

The next sum we consider is
\bea
T_2 =\sum_{\substack{\psi_1\in S_{N_{\hat A}}\times S_{N_{\hat B}}\\ \psi_2\in S_{N_{\hat A}'}\times S_{N_{\hat B}'}}}
\langle\vec{N}_{\hat A}',\vec{N}_{\hat B}'|\sigma_2\psi_2^{-1}E^{(1_{\hat A})}_{ci}E^{(1_{\hat B})}_{kc}\psi_1
|\vec{N}_{\hat A},\vec{N}_{\hat B}\rangle
\langle \vec{N}_{\hat A},\vec{N}_{\hat B}|\sigma_1^{-1}\psi_1^{-1}E^{(1_{\hat A})}_{ak}E^{(1_{\hat B})}_{ia}\psi_2
|\vec{N}_{\hat A}',\vec{N}_{\hat B}'\rangle
\nonumber
\eea
Changing variables $\psi_2^{-1}\to\tilde\psi_2^{-1}$  with
\bea
\tilde\psi_2^{-1}=\sigma_2\psi_2^{-1}\qquad\Rightarrow\qquad\tilde\psi_2 =\psi_2\sigma_2^{-1}
\eea
the sum becomes
\bea
T_2 &=&
\sum_{\substack{\psi_1\in S_{N_{\hat A}}\times S_{N_{\hat B}}\\ \psi_2\in S_{N_{\hat A}'}\times S_{N_{\hat B}'}}}
\langle\vec{N}_{\hat A}',\vec{N}_{\hat B}'|\psi_2^{-1}E^{(1_{\hat A})}_{ci}E^{(1_{\hat B})}_{kc}\psi_1
|\vec{N}_{\hat A},\vec{N}_{\hat B}\rangle\cr
&&\times\langle \vec{N}_{\hat A},\vec{N}_{\hat B}|\sigma_1^{-1}\psi_1^{-1}E^{(1_{\hat A})}_{ak}
E^{(1_{\hat B})}_{ia}\psi_2\sigma_2 |\vec{N}_{\hat A}',\vec{N}_{\hat B}'\rangle\cr\cr
&=&\sum_{\substack{\psi_1\in S_{N_{\hat A}}\times S_{N_{\hat B}}\\ \psi_2\in S_{N_{\hat A}'}\times S_{N_{\hat B}'}}}
\langle\vec{N}_{\hat A}',\vec{N}_{\hat B}'|\psi_2^{-1}\psi_1E^{\psi_1^{-1}(1_{\hat A})}_{ci}
E^{\psi_1^{-1}(1_{\hat B})}_{kc}|\vec{N}_{\hat A},\vec{N}_{\hat B}\rangle\cr
&&\times\langle \vec{N}_{\hat A},\vec{N}_{\hat B}|\sigma_1^{-1}E^{\psi_1^{-1}(1_{\hat A})}_{ak}
E^{\psi_1^{-1}(1_{\hat B})}_{ia}\psi_1^{-1}\psi_2\sigma_2 |\vec{N}_{\hat A}',\vec{N}_{\hat B}'\rangle
\nonumber
\eea
Change variables $\psi_2\to\rho$ with $\rho=\psi_1^{-1}\psi_2$ and relabel $\rho\to\psi_2$ to find
\bea
T_2 &=&\sum_{\substack{\psi_1\in S_{N_{\hat A}}\times S_{N_{\hat B}}\\ \psi_2\in S_{N_{\hat A}'}\times S_{N_{\hat B}'}}}
\langle\vec{N}_{\hat A}',\vec{N}_{\hat B}'|\psi_2^{-1}E^{\psi_1^{-1}(1_{\hat A})}_{ci}E^{\psi_1^{-1}(1_{\hat B})}_{kc}
|\vec{N}_{\hat A},\vec{N}_{\hat B}\rangle\cr\cr
&&\times \langle \vec{N}_{\hat A},\vec{N}_{\hat B}|\sigma_1^{-1}E^{\psi_1^{-1}(1_{\hat A})}_{ak}
E^{\psi_1^{-1}(1_{\hat B})}_{ia}\psi_2\sigma_2 |\vec{N}_{\hat A}',\vec{N}_{\hat B}'\rangle
\nonumber
\eea
Recall that $\vec{v}_b$ denotes the $p$ dimensional vector with all entries zero except the $b$th entry, which is 1.
For a non-zero contribution, the factor on the first line above requires that
\bea
\vec{N}_{\hat A}-\vec{v}_i+\vec{v}_c &=& \vec{N}_{\hat A}'\cr
\vec{N}_{\hat B}-\vec{v}_c+\vec{v}_k &=& \vec{N}_{\hat B}'
\eea
and the factor on the second line above requires
\bea
\vec{N}_{\hat B}-\vec{v}_i+\vec{v}_a &=&\vec{N}_{\hat B}'\cr
\vec{N}_{\hat A}-\vec{v}_a+\vec{v}_k &=& \vec{N}_{\hat A}'
\eea
There are two solutions:

{\vskip 0.25cm}

\underline{Case 1:} $\vec{v}_c=\vec{v}_i$ and $\vec{v}_a=\vec{v}_k$.
In this case $\vec{N}_{\hat A}=\vec{N}_{\hat A}'$  and $\vec{N}_{\hat B}-\vec{v}_i+\vec{v}_k =\vec{N}_{\hat B}'$.

\underline{Case 2:} $\vec{v}_c=\vec{v}_k$ and $\vec{v}_a =\vec{v}_i$.
In this case $\vec{N}_{\hat B}=\vec{N}_{\hat B}'$  and $\vec{N}_{\hat A}-\vec{v}_i+\vec{v}_k =\vec{N}_{\hat A}'$.

{\vskip 0.25cm}

\noindent
The analysis for case 1 is as follows
\bea
T_2 &=&\sum_{\substack{\psi_1\in S_{N_{\hat A}}\times S_{N_{\hat B}}\\ \psi_2\in S_{N_{\hat A}}\times S_{N_{\hat B}'}}}
\langle\vec{N}_{\hat A},\vec{N}_{\hat B}'|\psi_2^{-1}E^{\psi_1^{-1}(1_{\hat A})}_{ii}
E^{\psi_1^{-1}(1_{\hat B})}_{ki}|\vec{N}_{\hat A},\vec{N}_{\hat B}\rangle\cr\cr
&&\qquad\qquad\qquad\qquad
\times \langle \vec{N}_{\hat A},\vec{N}_{\hat B}|\sigma_1^{-1}E^{\psi_1^{-1}(1_{\hat A})}_{kk}
E^{\psi_1^{-1}(1_{\hat B})}_{ik}\psi_2\sigma_2 |\vec{N}_{\hat A},\vec{N}_{\hat B}'\rangle
\label{enlightening}\cr\cr
&=& (N_{\hat A} - 1)!(N_{\hat B} - 1)!(N_{\hat A})_{i\to k}(N_{\hat B})_{ii}\cr\cr
&&\qquad\qquad\qquad\qquad\times
\sum_{\psi_2\in S_{N_{\hat A}}\times S_{N_{\hat B}'}}
\langle\vec{N}_{\hat A},\vec{N}_{\hat B}'|\psi_2^{-1}|\vec{N}_{\hat A},\vec{N}_{\hat B}'\rangle
\langle \vec{N}_{\hat A},\vec{N}_{\hat B}'|\sigma_1^{-1}\psi_2\sigma_2 |\vec{N}_{\hat A},\vec{N}_{\hat B}'\rangle\cr\cr
&=& (N_{\hat A} - 1)!(N_{\hat B} - 1)!(N_{\hat A})_{i\to k}(N_{\hat B})_{ii}
\sum_{\psi_2\in S_{N_{\hat A}}\times S_{N_{\hat B}'}}\sum_{h_1,h_2\in H_2}
\delta (\psi_2^{-1}h_1)\delta (\sigma_1^{-1}\psi_2\sigma_2h_2)\cr\cr
&=& (N_{\hat A}- 1)!(N_{\hat B} - 1)!(N_{\hat A})_{i\to k}(N_{\hat B})_{ii}\sum_{h_1,h_2\in H_2}
\delta (\sigma_1^{-1}h_1\sigma_2h_2)
\eea
A few comments are in order.
The operator $E^{\psi_1^{-1}(1_{\hat A})}_{ii}$ acts directly on $|\vec{N}_{\hat A}\rangle$ in the first line above, while
the operator $E^{\psi_1^{-1}(1_A)}_{kk}$ acts on $\sigma_1|\vec A\rangle$ in the second line.
This is only non-zero for strings stretching from $i$ to $k$.
The factor $E^{\psi_1^{-1}(1_B)}_{ki}$ removes a closed loop from node $i$ in $\sigma_1$ and moves it
to node $k$ in $\sigma_2$.
This term allows edges that have both endpoints on a single node to hop between nodes.
Note that $\vec{N}_{\hat A}=\vec{N}_{\hat A}'$, but $\vec{N}_{\hat B}\ne \vec{N}_{\hat B}'$.
The final inner products are for vectors $|\vec{N}_{\hat A}',\vec{N}_{\hat B}'\rangle$ and that is why we land up
summing over $H_2$.
If we first do the sum over $\psi_2$ and then the sum over $\psi_1$, we find that
\bea
T_2 = (N_{\hat A}- 1)!(N_{\hat B} - 1)!(N_{\hat A}')_{k\to i}(N_{\hat B}')_{kk}
\sum_{h_1,h_2\in H_1}
\delta (\sigma_1^{-1}h_1\sigma_2h_2)\label{secondform}
\eea
It is straightforward to verify the equivalence of (\ref{enlightening}) and (\ref{secondform}).
We now turn to the analysis for case 2.
The analysis proceeds along the same lines as for case 1.
The sum we need to perform is
\bea
T_2 &=&\sum_{\substack{\psi_1\in S_{N_{\hat A}}\times S_{N_{\hat B}}\\ \psi_2\in S_{N_{\hat A}'}\times S_{N_{\hat B}}}}
\langle\vec{N}_{\hat A}',\vec{N}_{\hat B}|\psi_2^{-1}E^{\psi_1^{-1}(1_{\hat A})}_{ki}
E^{\psi_1^{-1}(1_{\hat B})}_{kk}|\vec{N}_{\hat A},\vec{N}_{\hat B}\rangle\cr\cr
&&\qquad\qquad\qquad\qquad
\times\langle \vec{N}_{\hat A},\vec{N}_{\hat B}|\sigma_1^{-1}E^{\psi_1^{-1}(1_{\hat A})}_{ik}
E^{\psi_1^{-1}(1_{\hat B})}_{ii}\psi_2\sigma_2 |\vec{N}_{\hat A}',\vec{N}_{\hat B}\rangle\cr\cr
&=& (N_{\hat A} - 1)!(N_{\hat B} - 1)!(N_{\hat A})_{ii}(N_{\hat B})_{k\to i}\cr\cr
&&\qquad\qquad\qquad\qquad\sum_{\psi_2\in S_{N_{\hat A}'}\times S_{N_{\hat B}}}
\langle\vec{N}_{\hat A}',\vec{N}_{\hat B}|\psi_2^{-1}|\vec{N}_{\hat A},\vec{N}_{\hat B}'\rangle
\langle\vec{N}_{\hat A}',\vec{N}_{\hat B}|\sigma_1^{-1}\psi_2\sigma_2 |\vec{N}_{\hat A}',\vec{N}_{\hat B}\rangle\cr\cr
&=& (N_{\hat A} - 1)!(N_{\hat B} - 1)!(N_{\hat A})_{ii}(N_{\hat B})_{k\to i}
\sum_{\psi_2\in S_{N_{\hat A}'}\times S_{N_{\hat B}}}
\sum_{h_1,h_2\in H_{\vec{N}_{\hat A}'}\times H_{\vec{N}_{\hat B}}}
\delta (\psi_2^{-1}h_1)\delta (\sigma_1^{-1}\psi_2\sigma_2h_2)\cr\cr
&=& (N_{\hat A} - 1)!(N_{\hat B} - 1)!(N_{\hat A})_{ii}(N_{\hat B})_{k\to i}
\sum_{h_1,h_2\in H_2}\delta (\sigma_1^{-1}h_1\sigma_2h_2)
\eea
Thus, for $T_2$ we find
\bea
T_2 = (N_{\hat A}-1)!(N_{\hat B} - 1)!\Big((N_{\hat A})_{i\to k}(N_{\hat B})_{ii}
+(N_{\hat A})_{ii}(N_{\hat B})_{k\to i}\Big)
\sum_{h_1,h_2\in H_2}\delta (\sigma_1^{-1}h_1\sigma_2h_2)\cr
\eea
which can also be written as
\bea
T_2 = (N_{\hat A} - 1)!(N_{\hat B} - 1)!\Big((N_{\hat A}')_{k\to i}(N_{\hat B}')_{kk}
+(N_{\hat A}')_{kk}(N_{\hat B}')_{i\to k}\Big)
\sum_{h_1,h_2\in H_1}\delta (\sigma_1^{-1}h_1\sigma_2h_2)\nonumber
\eea
A very similar analysis now gives
\bea
T_3 = (N_{\hat A} - 1)!(N_{\hat B} - 1)!\Big((N_{\hat A})_{k\to i}(N_{\hat B})_{ii}
+(N_{\hat A})_{ii}(N_{\hat B})_{i\to k}\Big)
\sum_{h_1,h_2\in H_2}\delta (\sigma_1^{-1}h_1\sigma_2h_2)\cr
\eea
which can also be written as
\bea
T_3 = (N_{\hat A} - 1)!(N_{\hat B} - 1)!\Big((N_{\hat A}')_{i\to k}(N_{\hat B}')_{kk}
+(N_{\hat A}')_{kk}(N_{\hat B}')_{k\to i}\Big)
\sum_{h_1,h_2\in H_1}\delta (\sigma_1^{-1}h_1\sigma_2h_2)\cr
\eea
Summing the four contributions, we now obtain a rather simple formula for the matrix elements
\bea
(M_{\hat A\hat B})_{R,r_1,\vec\sigma_1\,\,T,t_1,\vec\sigma_2}=
\prod_{\hat C\ne\hat A,\hat B}\delta_{(\sigma_{\hat C})_1(\sigma_{\hat C})_2} \sum_{R'}
{\delta_{r_1t_1}\delta_{R'_iT'_k}\over\sqrt{|O_{R,r_1}(\vec\sigma_1)|^2 |O_{T,t_1}(\vec\sigma_2)|^2}}
\sqrt{c_{RR'}c_{TT'}\over l_{R_i}l_{T_k}}\cr
\times [2\delta_{ik}(N_{\hat A})_i (N_{\hat B})_i -\Big((N_{\hat A})_{ki}(N_{\hat B})_{ii}+(N_{\hat A})_{ii}(N_{\hat B})_{ik}\Big)]
\sum_{h_1,h_2\in H_2}
\delta (\sigma_1^{-1}h_1\sigma_2h_2)\cr
\label{finalDgen}
\eea
which can also be written as
\bea
(M_{\hat A\hat B})_{R,r_1,\vec\sigma_1\,\,T,t_1,\vec\sigma_2}=
\prod_{\hat C\ne\hat A,\hat B}\delta_{(\sigma_{\hat C})_1(\sigma_{\hat C})_2}\sum_{R'}
{\delta_{r_1t_1}\delta_{R'_iT'_k}\over\sqrt{|O_{R,r_1}(\vec\sigma_1)|^2 |O_{T,t_1}(\vec\sigma_2)|^2}}
\sqrt{c_{RR'}c_{TT'}\over l_{R_i}l_{T_k}}\cr
\times [2\delta_{ik}(N_{\hat A}')_i (N_{\hat B}')_i -\Big((N_{\hat A}')_{ki}(N_{\hat B}')_{kk}+(N_{\hat A}')_{kk}
(N_{\hat B}')_{ik}\Big)]\sum_{h_1,h_2\in H_1}\delta (\sigma_1^{-1}h_1\sigma_2h_2)\cr
\label{LastM}
\eea
Finally, note that the norm of the Gauss graph operator is given by
\bea
|O_{R,r_1}(\vec\sigma)|^2
=\prod_{\hat A=1}^4 \prod_{i,j=1}^{p} (N^{\vec\sigma}_{\hat A})_{i\to j}!
\eea
The result (\ref{LastM}) is one of the new results of this paper.

\section{Emergent Lattice Model}\label{EmergentLattice}

The formula for the dilatation operator in Gauss graph basis has a fascinating structure.
There are two types of terms.
There are four terms mixing $\phi_1$ with the excitations, summarized in (\ref{ggaction}).
These terms do not act on the Gauss graph label i.e. operators that mix have the same Gauss graph
label, but different $R,r_1$ labels.
There are also six terms, mixing the excitations, summarized in (\ref{exdil}) and (\ref{finalDgen}).
These terms act only on the Gauss graph labels i.e. operators that mix have the same $R,r_1$ labels,
but different Gauss graph labels.
Consequently, these terms can be simultaneously  diagonalized.

The mixing between operators with different Gauss graphs is tightly constrained.
Recall that edges in the Gauss graph come in four species, one for each excitation $\hat A$.
The edges are oriented and the number of edges of each species entering each node must match the
number of edges of the same species leaving the node.
The dilatation operator only mixes Gauss graphs that have the same number of edges of each species.
There is an even tighter constraint on the mixing: graphs can only mix if they have exactly the same number,
orientation and species of edges stretching between distinct nodes.
Consequently, if two operators mix their graphs differ only by the placement of the edges that have both endpoints
attached to a single node.

In this section we would like to interpret the dilatation operator as a Hamiltonian acting on the Gauss graph, using ideas
first described in \cite{deCarvalho:2018xwx}.
The dynamics is all in the closed edges that have both end points attached to a node.
We will identify these closed edges as particles hopping on a lattice, with lattice sites given by the nodes of the
Gauss graph.
Of course, each node in the Gauss graph corresponds to a row in $R$, and each row in $R$ corresponds to a giant
graviton brane.
In the next section we will show that these closed edges are in fact quanta of the brane worldvolume theory.
To obtain the ``graph dynamics'' we introduce a collection of creation and annihilation operators, one for each species of edge.
The matrix elements (\ref{ggaction}) and (\ref{finalDgen}) are written entirely in terms of the number of edges appearing
in the graph.
If we translate each graph into a Fock space state, by interpreting the graph as an occupation number representation of
the state, then the number of edges can be written using the usual number operator.
The hopping of closed edges between nodes is easily accomplished by destroying an edge at one node and creating it
at another.

To proceed, introduce two sets of bosonic oscillator operators, $(b_1)_{ij},(\bar b_1)_{ij}$ for $\phi_2$ corresponding
to $\hat A=1$ and  $(b_2)_{ij},(\bar{b}_2)_{ij}$ for $\phi_3$ ($\hat A=2$), as well as two sets of fermionic
oscillator operators,
$(f_1)_{ij},(\bar f_1)_{ij}$ for $\psi_1$ ($\hat A=3$) and $(f_2)_{ij},(\bar{f}_2)_{ij}$ for $\psi_2$ ($\hat A=4$).
Since we want to create and destroy edges with end points at any two nodes, the indices $i,j$ must range over $1,2,\cdots,p$.
Thus, the dynamics is that of $p\times p$ matrices, where we recall that the number of rows in $R$ is $p$.
Note that the original theory is based on a gauge theory with U$(N)$ gauge group and hence it involves $N\times N$ and not
$p\times p$ matrices.
To refer to the complete collection of bosonic and fermionic oscillators we will use $(a_{\hat A})_{ij},(\bar a_{\hat A})_{ij}$.
The oscillator algebra is ($a,b=1,2$)
\bea
\big[ (b_a)_{ij},(\bar{b}_b)_{kl}\big]&=&\delta_{ab}\delta_{il}\delta_{jk}\qquad\qquad
\big[ (b_a)_{ij},(b_b)_{kl}\big]=\big[ (\bar{b}_a)_{ij},(\bar{b}_b)_{kl}\big]=0\cr\cr
\big\{ (f_a)_{ij},(\bar{f}_b)_{kl}\big\}&=&\delta_{ab}\delta_{il}\delta_{jk}\qquad\qquad
\big\{ (f_a)_{ij},(f_b)_{kl}\big\}=\big\{ (\bar{f}_a)_{ij},(\bar{f}_b)_{kl}\big\}=0
\eea
The vacuum of Fock space $|0\rangle$ obeys $(b_a)_{ij}|0\rangle =0=(f_a)_{ij}|0\rangle$ for $i,j = 1,2,\cdots ,p$.
The Gauss graph operators are now represented as states in Fock space, as follows
\bea
O_{R,r}(\vec\sigma)\quad\longleftrightarrow\quad
\prod_{\hat A=1}^4 \prod_{i,j=1}^p (\bar a_{\hat A})_{ij}^{(N_{\hat A})_{i\to j}}|0\rangle
\eea

Out next task is to represent the dilatation operator $D_{\hat A\hat B}$ in the Gauss graph basis.
The product of delta functions appearing in (\ref{finalDgen}) is not normalized.
We will trade it for a delta function normalized to 1
\bea
\prod_{\hat C\ne\hat A,\hat B}\delta_{(\sigma_{\hat C})_1(\sigma_{\hat C})_2}
\delta (\sigma_1^{-1}h_1\sigma_2h_2) = |O_{R,r_1}(\vec\sigma_1)|^2 \delta_{[\vec\sigma_1][\vec\sigma_2]}
\eea
where $\delta_{[\vec\sigma_1][\vec\sigma_2]}=1$ if permutations $\vec\sigma_1$ and $\vec\sigma_2$ belong to the
same class of the cosets (\ref{dcosets}), and the delta function vanishes if they are not in the same class.
The matrix elements $(M_{\hat A\hat B})_{R,r_1,\vec\sigma_1\,\,T,t_1,\vec\sigma_2}$ are only non-zero if we can choose coset
representatives such that $\sigma_1$ and $\sigma_2$ describe the same element of $S_{N_{\hat A}}\times S_{N_{\hat B}}$.
This reflects the fact that the graphs described by $\sigma_1$ and $\sigma_2$ differ only in the number of edges with both
ends attached to the same node, but not in the number of edges between distinct nodes.
In this case the matrix element in (\ref{finalDgen}) simplifies to
\bea
(M_{\hat A\hat B})_{R,r_1,\vec\sigma_1\,\,T,t_1,\vec\sigma_2}&=&
\sum_{R'}{\sqrt{|O_{R,r_1}(\sigma_1)|^2\over |O_{T,t_1}(\sigma_2)|^2}}
\delta_{r_1t_1}\delta_{R'_iT'_k}\delta_{[\vec\sigma_1][\vec\sigma_2]}
\sqrt{(N+l_{R_i})(N+l_{T_k})\over l_{R_i}l_{T_k}}\cr\cr
&\times& \Big[2\delta_{ik}(N_{\hat A})_i (N_{\hat B})_i -
\Big((N_{\hat A})_{ki}(N_{\hat B})_{kk}+(N_{\hat A})_{kk}(N_{\hat B})_{ik}\Big)\Big]
\eea
Using the oscillators introduced above, we can write number operators whose eigenvalues count the edges in the graph.
We will use a hat when we want to describe a number operator which acts on states and no hat when we want to refer to
the integer number of edges of a particular graph.
For example
\bea
(\hat N_{\hat A})_{ii} = (\bar a_{\hat A})_{ii} (a_{\hat A})_{ii}\qquad
(\hat N_{\hat A})_{i\to k} = (\bar a_{\hat A})_{ki} (a_{\hat A})_{ik}
\eea
where there is no sum on $i,k$ in the last formula above, and
\bea
(N_{\hat A})_i &=& \sum_{k\ne i} (\hat{N}_{\hat A})_{i\to k}
+(\bar{a}_{\hat A})_{ii} (a_{\hat A})_{ii}=\sum_{k} (\bar{a}_{\hat A})_{ki} (a_{\hat A})_{ik}\cr
&=& \sum_{k\ne i} (\hat{N}_{\hat A})_{k\to i}+(\bar{a}_{\hat A})_{ii} (a_{\hat A})_{ii}
=\sum_{k\ne i} (\bar{a}_{\hat A})_{ik} (a_{\hat A})_{ki}
\eea
We can then write the piece of the Hamiltonian of the lattice model we are considering as
\bea
H_{\hat A\hat B} &=&\sum_{i,j=1}^p\sqrt{(N+l_{R_i})(N+l_{R_j})\over l_{R_i}l_{R_j}}\Bigg(
-(\hat{N}_{\hat B})_{ji}(\bar{a}_{\hat A})_{jj} (a_{\hat A})_{ii}
-(\hat{N}_{\hat A})_{ji}(\bar{a}_{\hat B})_{jj} (a_{\hat B})_{ii}\cr\cr
&&\quad +2\delta_{ij}
\Big(\sum_{l\ne i} (\hat{N}_{\hat A})_{i\to l} + (\bar{a}_{\hat A})_{ii}(a_{\hat A})_{ii}\Big)
\Big(\sum_{l\ne i} (\hat{N}_{\hat B})_{i\to l} + (\bar{a}_{\hat B})_{ii}(a_{\hat B})_{ii}\Big)
\Bigg)
\label{LatticeAB}
\eea
The complete Hamiltonian is obtained by summing over $A,B$.
Matrix elements of (\ref{LatticeAB}) computed using the Fock space states are in exact agreement with matrix elements of
the one loop dilatation operator, computed in the Gauss graph basis.
Thus, our final result for the Hamiltonian of the lattice model, arising from the one loop dilation operator, is
\bea
H&=&-{2g_{YM}^2\over (4\pi)^2}\sum_{\hat A=1}^4\sum_{i>j=1}^p(\hat{N}_{\hat A})_{ij}\Delta_{ij}\cr
&&-{2g_{YM}^2\over (4\pi)^2}\sum_{\hat{A}=1}^3\sum_{\hat B=1+\hat A}^4
\sum_{i,j=1}^p\sqrt{(N+l_{R_i})(N+l_{R_j})\over l_{R_i}l_{R_j}}\Bigg(
-(\hat{N}_{\hat B})_{ji}(\bar{a}_{\hat A})_{jj} (a_{\hat A})_{ii}
-(\hat{N}_{\hat A})_{ji}(\bar{a}_{\hat B})_{jj} (a_{\hat B})_{ii}\cr\cr
&&\quad +2\delta_{ij}
\Big(\sum_{l\ne i} (\hat{N}_{\hat A})_{i\to l} + (\bar{a}_{\hat A})_{ii}(a_{\hat A})_{ii}\Big)
\Big(\sum_{l\ne i} (\hat{N}_{\hat B})_{i\to l} + (\bar{a}_{\hat B})_{ii}(a_{\hat B})_{ii}\Big)
\Bigg)
\label{LatticeH}
\eea

\section{Emergent Yang-Mills Theory}\label{EYM}

The operators we study are labeled by Young diagrams that have $p$ long rows.
They are holographically dual to a system of $p$ dual giant gravitons that have expanded to S$^3\subset$AdS$_5$.
A natural guess is that the dynamics described by the Hamiltonian we have derived arises from the worldvolume
dynamics of a system of $p$ giant gravitons.
In this section we will confirm this expectation.

This worldvolume theory of the giant graviton branes comes from the dynamics of their open string excitations, so we expect
the world volume dynamics is a super Yang-Mills theory.
Since the space defined by the brane's world volume is not the space on which the original gauge theory is defined, we will
refer to this as an emergent gauge theory\cite{Balasubramanian:2004nb}.
We can say a few things about precisely what theory we expect:
\begin{itemize}
\item[1.] Since there are $p$ giant graviton branes we expect a U$(p)$ gauge theory.
Each brane corresponds to a row in the Young diagram, and therefore, to a node in the Gauss graph.
The edges which stretch between (not necessarily distinct) nodes will be identified with the open string excitations.
Thus, if we label the nodes in the graph with an integer $i=1,2,\cdots,p$, we naturally label the end points of the
edges by allowing them to inherit the label of the node.
These labels for the end points of the edges are the Chan-Paton indices of the open strings.
\item[2.] Before adding any excitations, the operators are constructed from a single field $\phi_1$ and are 1/2 BPS.
The brane moves in AdS spacetime with metric
\bea
ds^2 =R^2 (-\cosh ^{2}\rho \,dt^2+\,d\rho ^{2}+\sinh ^{2}\rho \,d\Omega_3^2)
\eea
The $\rho$ at which the giant is located is specified by
\bea
\cosh\rho = \sqrt{1+{l_R\over N}}\qquad \sinh\rho =\sqrt{l_R\over N}
\eea
with $l_R$ the length of the row in $R$ corresponding to the giant graviton brane.
We are in the displaced corners approximation, which implies that the row lengths of our operators are unequal and hence
the $p$-branes are separated in spacetime.
Consequently we are studying the gauge theory on its Coulomb branch.
In the low energy limit the dynamics is described by a U$(1)^p$ gauge theory.
This nicely matches what we find: our dynamical fields are the closed loops formed by edges located at a given node - these
are the only edges that are changed by the action of the dilatation operator.
The open strings corresponding to these dynamical edges have both end points labeled by the same gauge group index, so
they belong to the diagonal U$(1)^p$.
There is one U$(1)$ for each node.
Notice that $g_{YM}^2$ of the emergent theory is equal to the AdS$_5\times$S$^5$ string coupling which is itself equal
to the original coupling $g_{YM}^2$ of the ${\cal N}=4$ super Yang-Mills theory we study.
Since we are studying weak coupling in the original Yang-Mills theory, we are at weak coupling in the emergent gauge theory.
\item[3.] We have not studied the complete ${\cal N}=4$ super Yang-Mills theory, since we have truncated to the su$(2|3)$
sector.
Consequently, we will only recover part of the expected U$(1)^p$ gauge theory.
The bosonic part of the symmetry of ${\cal N}=4$ super Yang-Mills theory is SO$(2,4)\times$SO$(6)$.
The SO$(4)$ that acts as the isometry of the brane world volume is a subgroup of SO$(2,4)$; this SO$(4)$ is a spacetime
symmetry of the world volume theory.
The SO$(4)$ which rotates the real components of the $\phi_2,\phi_3$ fields is a subgroup of SO$(6)$; this SO$(4)$ is a
global symmetry of the world volume theory.
Consequently the excitations constructed from the $\phi_2,\phi_3$ fields are scalar fields of the emergent gauge theory.
$\psi_1$ and $\psi_2$ are their super partners.
Finally, our truncation to the su$(2|3)$ sector retains only fields invariant under the SO$(4)\subset$SO$(4,2)$, so that
we should expect to reproduce the s-wave sector of the emergent gauge theory.
\end{itemize}
Thus, we should compare our emergent theory to the low energy limit of a U$(p)$ gauge theory on its Coulomb branch.
We expect to reproduce the $s$-wave sector of the dynamics of the adjoint scalars and their super partners.

With these comments in mind, we now consider the action for the adjoint scalars of a U$(p)$ gauge theory, defined on an
S$^3$, which has the form
\bea
S&=&{1\over g_{YM}^2}\int_{\mathbb{R}\times S^3}\Big[
{\rm Tr}\Big(
\partial_\mu X\partial^\mu X^\dagger + \partial_\mu Y\partial^\mu Y^\dagger
-{1\over R^2}(XX^\dagger+YY^\dagger)
-\, [X,Y][Y^\dagger,X^\dagger]\Big)\cr
&&\qquad\qquad-\sum_{i\ne j}m_{ij}^2(X_{ij}X^\dagger_{ji}+Y_{ij}Y^\dagger_{ji})
\Big]dt\, R^3 d\Omega_3
\eea
$R$ is the radius of the S$^3$ on which the theory is defined and the $1/R^2$ terms are required for conformal invariance,
as usual.
The off diagonal matrix elements of adjoint scalars $X,Y$ will have masses $m_{ij}$ proportional to the distances separating
the branes between which they stretch.
Truncating to the $s$-wave sector gives the matrix quantum mechanics
\bea
S&=&{R^3\Omega_3\over g_{YM}^2}\int_{\mathbb{R}} \Big[{\rm Tr}\Big(
\dot{X}\dot{X}^\dagger + \dot{Y}\dot{Y}^\dagger
-{1\over R^2}(XX^\dagger+YY^\dagger)
-\, [X,Y][Y^\dagger,X^\dagger]\Big)\cr
&&\qquad\qquad-\sum_{i\ne j}m_{ij}^2(X_{ij}X^\dagger_{ji}+Y_{ij}Y^\dagger_{ji})dt
\Big]
\eea
The eigenvalues of the one loop dilatation operator give the spectrum of anomalous dimensions.
Identifying the classical contribution to the dimension with the free part of the emergent gauge theory,
the dynamics obtained from the one loop dilatation operator should match the interaction Hamiltonian, given by
\bea
H_{\rm int}={R^3\Omega_3\over g_{YM}^2}\left[\sum_{i\ne j}m_{ij}^2(X_{ij}X^\dagger_{ji}+Y_{ij}Y^\dagger_{ji})
+{\rm Tr}\left([X,Y][Y^\dagger,X^\dagger]\right)\right]
\eea
The operators studied in earlier sections are constructed using $\phi_2,\phi_3$ and not $\phi_2^\dagger,\phi_3^\dagger$.
This truncation must also be accounted for.
This is achieved by truncating the mode expansions
\bea
X={1\over\sqrt{2}}(\tilde a+ \bar{a})\qquad\qquad
X^\dagger={1\over\sqrt{2}}( a+\bar{\tilde{a}})\cr
Y={1\over\sqrt{2}}(\tilde b+ \bar{b})\qquad\qquad
Y^\dagger={1\over\sqrt{2}}( b+\bar{\tilde{b}})
\eea
The truncation sets all tilded oscillators to zero, i.e. we replace $X\to \bar{a}$, $X^\dagger\to a$, $Y\to\bar{b}$ and
$Y^\dagger\to b$.
The interaction Hamiltonian becomes (we are assuming normal ordering for $H_{\rm int}$)
\bea
H_{\rm int}={R^3\Omega_3\over g_{YM}^2}\left[\sum_{i\ne j}m_{ij}^2(\bar{a}_{ij}a_{ji}+\bar{b}_{ij}b_{ji})
+{\rm Tr}\left([\bar{b},\bar{a}][a,b]\right)\right]\label{toreproduce}
\eea
The first term in the interaction Hamiltonian matches the terms in the dilatation operator with action given in (\ref{ggaction}).
To see the equality, note that at large $N$ we are justified in ignoring the difference between
$O_{R^+_{ij},(r^+_{ij},s)\mu_1\mu_2}$ and $O_{R,(r,s)\mu_1\mu_2}$, which amounts to ignoring the effects of
back reaction, due to the open string excitations, on the size of the giants.
Once the back reaction is ignored, the action quoted in (\ref{ggaction}) simplifies nicely.
For example \footnote{There is a similar result for all $D_{\phi_1\hat A}$ terms.}
\bea
D_{\phi_1\phi_2}|\sigma\rangle =-\sum_{i,j=1}^p
\Big(\sqrt{N+l_{R_i}}-\sqrt{N+l_{R_j}}\Big)^2\bar a_{ij}a_{ij}|\sigma\rangle
\label{sqrddist}
\eea
How should we interpret this answer?
Our giant gravitons are constructed mainly from $\phi_1$ fields, with a small number of excitations.
Consequently, they are small deformations of ${1\over 2}$ BPS operators.
A very natural set of coordinates for the study of ${1\over 2}$ BPS geometries in the dual gravitational theory was
given by Lin, Lunin and Maldacena in \cite{LLM}.
The geometry is written in terms of two three spheres, time $t$ and three more spacial coordinates $y,x_1,x_2$.
In terms of these coordinates, the AdS$_5\times$S$^5$ geometry corresponds to a circular droplet boundary condition
on the $y=0$ plane, parameterized by the $(x_1,x_2)$ coordinates (see section 2.3 of \cite{LLM}).
Introduce radial coordinates $(r,\phi)$ on this plane.
The $r$ and $y$ coordinates are related to $\rho$ (the radial variable of AdS$_5$ in global coordinates) and
$\theta$ (one of the angles of the S$^5$) by $y=r_0\sinh\rho\sin\theta$ and $r=r_0\cosh\rho\cos\theta$,
where $r_0=R_{{\rm AdS}_5}^2=R_{{\rm S}^5}^2$.
The dual giant gravitons corresponding to a row of length $l_R$ is located at
\bea
 \theta=0\qquad\cosh\rho=\sqrt{1+{l_R\over N}}
\eea
so that
\bea
y=0 \qquad r=\sqrt{1+{l_R\over N}}
\eea
From the AdS$_5\times$S$^5$ geometry written in LLM coordinates, we find that the metric on the LLM plane at $y=0$
is given by $ds^2=(dx_1)^2+(dx_2)^2=dr^2+r^2d\phi^2$.
Thus, the coefficient in (\ref{sqrddist}) is square of the proper distance between the branes corresponding to
rows $i$ and $j$ of $R$.
This proves that (\ref{ggaction}) reproduces the first term in (\ref{toreproduce}) after identifying $a_{ij},\bar{a}_{ij}$ with
$(b_1)_{ij},(\bar{b}_1)_{ij}$.
Notice further that the squared masses are indeed proportional to the square of distances between branes.
In the same way, the oscillators $b_{ij},\bar{b}_{ij}$ will produce the required mass terms for $(b_2)_{ij},(\bar{b}_2)_{ij}$.

Now consider the commutator squared term
\bea
{\rm Tr}\left( [\bar b,\bar a][a,b]\right)&=&{\rm Tr}(\bar b\bar aab+\bar a\bar bba-\bar a\bar bab-\bar b\bar aba)
\eea
We need to perform a truncation to obtain the low energy theory.
The truncation will freeze the dynamics of the massive modes.
This is most simply illustrated with a specific example: consider the term
${\rm Tr}(\bar{b}\bar{a}ab)=\bar{b}_{ij}\bar{a}_{jk}a_{kl}b_{li}$.
Borrowing the language of the Gauss graph to make the discussion transparent, this term destroys a $b$ edge stretching
from $i$ to $l$ and creates a $b$ edge stretching from $i$ to $j$.
To freeze the edges stretched between nodes we should keep only the terms with $j=l$.
Truncating to achieve this we find
\bea
{\rm Tr}(\bar b\bar aab)&=&\bar b_{ij}\bar a_{jk}a_{kl}b_{li}\cr
&\to&\bar b_{ij}\bar a_{jk}a_{kj}b_{ji}
=\bar b_{ij}b_{ji}\bar a_{jk}a_{kj}\cr
&=&\sum_j (\sum_{i\ne j} (\hat n_3)_{j\to i}+ (\hat n_3)_{jj})(\sum_{k\ne j} (\hat n_2)_{j\to k}+ (\hat n_2)_{jj})
\eea
This truncation can be viewed as a Born-Oppenheimer approximation, in which we fix the edges stretched between nodes
and solve the dynamics of the light edges.
This will be a good approximation as long as we don't excite the light edges to an energy comparable to that of the
stretched edges.
Truncating the remaining terms in the commutator squared, we find
\bea
{\rm Tr}(\bar a\bar bba)&\to&(\sum_{i\ne j} (\hat n_2)_{j\to i}+ (\hat n_2)_{jj})
(\sum_{k\ne j} (\hat n_3)_{j\to k}+ (\hat n_3)_{jj})\cr
-{\rm Tr}(\bar a\bar bab)&\to&-(\hat n_2)_{ij}\bar b_{jj} b_{ii}-(\hat n_3)_{kj}\bar a_{jj}a_{kk}
+(\hat n_2)_{ii}(\hat n_3)_{ii}\cr
-{\rm Tr}(\bar b\bar aba)&\to&-(\hat n_3)_{ji} \bar a_{jj}a_{ii}-(\hat n_2)_{kj}\bar b_{jj} b_{kk}
+(\hat n_2)_{ii}(\hat n_3)_{ii}
\eea
Summing the four terms above we reproduce (\ref{LatticeAB}) in complete detail, up to the overall factor.
The overall factor given by
\bea
\propto -g_{YM}^2\sqrt{(N+l_{R_i})(N+l_{R_j})\over l_{R_i}l_{R_j}}
\eea
is perfectly explained as the field redefinition needed to match the dual giant graviton solution to a BPS
classical solution of super Yang-Mills theory on ${\mathbb R}\times$S$^3$ \cite{Hashimoto:2000zp}.
See Appendix \ref{FR} for a detailed discussion.

\section{Conclusions and Outlook}\label{discuss}

In this article we have studied the operator mixing problem for operators dual to systems of excited dual giant graviton branes.
The description we have constructed has a number of interesting features.
The mixing problem is simply described using a basis labeled by a pair of Young diagrams $R$ and $r_1$ and a graph
$\vec\sigma$.
The Young diagram $r_1$ organizes the $\phi_1$ fields in the operator.
In the dual holographic theory, each row corresponds to a dual giant graviton.
The length of the rows of $r_1$ is equal to the number of $\phi_1$ fields used to construct the giant and this gives the
momentum and hence the size of the (square of the) dual giant graviton.
The Young diagram $R$ plays a very similar role, except that it includes the excitations in the description.
The graph specifies the state of the excitations.
Nodes of the graph correspond to the giant gravitons, while the excitations are represented as edges with end points
attached to the nodes.
The matrix elements of the dilatation operator are written in terms of the number of edges appearing in the graph.
Interpreting the edges as an occupation number representation, we have mapped each Gauss graph operator into a
Fock space state and we have mapped the dilatation operator into a Hamiltonian acting on this Fock space.
We have identified this description with the Fock space of the emergent gauge theory, realized as the giant world
volume theory.

One obvious extension of our results would be to relax the truncation to the su$(2|3)$ sector.
By including fields that are not invariant under the SO$(4)$ rotating the world volume we go beyond the
$s$-wave sector.
This would start to reconstruct the spatial dependence of the world volume theory and constructing this
aspect of the world volume theory maybe a useful toy model for the emergence of spacetime in general.
Including the gauge fields for example, would be straight forward given the results already obtained
in \cite{deMelloKoch:2011vn}.
This would already be a fascinating and non-trivial extension.
Including further types of excitations would increase the number of Young diagrams labels on the restricted
Schur polynomials, as well as increasing the number of species of edges in the Gauss graphs.

As the number of giant gravitons grows one enters into the regime where back reaction can't be ignored.
In the description developed here, increasing the number of giant gravitons implies the number of nodes in the graph grows.
When the number of nodes becomes of order $N$, back reaction becomes important.
In the 1/2-BPS sector for example, states of $N$ giant gravitons back react to produce the LLM geometries\cite{LLM}.
In this regime the operators we study correspond to new spacetime geometries and it is interesting to ask if
signatures of the gravitational dynamics are visible.
The out-of-time-order correlator (OTOC) provides a signal of a possible gravity dual.
Holographic computations which consider shock waves in black hole geometries, has led to a bound on the quantum
Lyapunov exponent, evaluated using thermal OTOCs \cite{Maldacena:2015waa}.
The black hole geometries saturate the bound, with the maximum value attributed to the red shift near the event
horizon of the black hole.
To compute the thermal average we must average over all of the states in the Fock space.
Since the numbers $(N_{\hat A})_{i\to l}$ with $i\ne l$ label the state, the sum over states can be written as a sum over
these integers.
These same numbers appear as parameters in the Hamiltonian so that we are naturally lead to study a model for
particles hopping on a lattice with of the order of $N$ sites, with quenched disorder and, for the generic state, hopping
can happen between any two sites in the lattice i.e. all sites are connected.
These look a lot like the SYK models \cite{Kit,SaYe} which are known to saturate the chaos bound\cite{Maldacena:2016hyu},
suggesting that the computation of the OTOC for the lattice model developed here would be interesting.
Of course, the regime in which we expect to get a weakly curved gravity description is the limit of large 't Hooft coupling
and our dynamics is only one loop.
Nevertheless, the fact that to understand large $N$ but non-planar limits of ${\cal N}=4$ super Yang-Mills theory naturally
leads to models with quenched disorder and all-to-all interactions between the different sites, is interesting.

Another direction worth pursuing concerns the global symmetry of the model.
The dynamics of magnons in the planar limit is tightly constrained by the su$(2|2)$ symmetry of the model in an
interesting way\cite{Beisert:2005tm}.
The magnon ``polarizations'' fill out the fundamental representation of a centrally extended su$(2|2)$ algebra, which
enlarges the original algebra by two central charges $P$ and $K$.
These two additional central charges are related to gauge transformations which act non-trivially on individual fields.
By requiring that they annihilate the total state, one returns to the original global su$(2|2)$ symmetry.
This construction has a number of far reaching consequences.
First, it proves that the total anomalous dimension is a sum of contributions, one from each magnon.
Second, the kinematics of the global symmetry completely fixes the $S$-matrix, up to an overall phase.
The operators we study in this article enjoy the same global symmetry.
Is there a similar analysis to be developed for the operators dual to excited giant graviton branes?
This question was first explored in \cite{Berenstein:2014zxa}.
Recall that the lightest string modes of a string stretching between two flat parallel and separated D-branes
fill out a massive short representation of the unbroken supersymmetry of the D-brane system.
These representations require a central charge extension of the unbroken supersymmetry algebra.
The additional central charge has a physical interpretation as an electric charge carried by the open string end-points so that
closed string states are not charged.
An important conclusion of \cite{Berenstein:2014zxa} is that this open string central charge is a limit
of the central charge extension of \cite{Beisert:2005tm,Beisert:2006qh}.
The question was reconsidered in \cite{deCarvalho:2020pdp} using the language of the Gauss graph operators.
In the Gauss graph language, the magnons are the edges in the Gauss graph.
The conclusion of \cite{deCarvalho:2020pdp} is that edges stretched between nodes of the Gauss graph do carry the central
charge, while edges living at a node are not charged.
The central extension again generates gauge transformations so that it again vanishes when acting on physical states which
are gauge invariant.
In the double coset setting the constraint enforced by the Gauss Law (discussed in Section \ref{GGBasis}) ensures that
the central extension vanishes.
In the emergent dynamics that we have constructed in this article, edges with both ends attached to a single node are
gauge invariant, which immediately forces the central charge $P$ and $K$ to vanish for these edges.
This prevents us from repeating the analysis of \cite{Beisert:2005tm,Beisert:2006qh} to learn about the spectrum of
anomalous dimensions and the $S$-matrix of two magnon scattering.
It remains an interesting exercise to determine the constraints implied by the global su$(2|2)$ symmetry.

{\vskip 0.5cm}

\noindent
\begin{centerline}
{\bf Acknowledgements}
\end{centerline}

We would like to thank David Berenstein, Antal Jevicki and Sanjaye Ramgoolam for penetrating insights that were
helpful in completing this project.
This work is supported by the South African Research Chairs Initiative of the Department of Science and Technology and
National Research Foundation of South Africa as well as by funds received from the National Institute for Theoretical
Physics (NITheP).

\appendix
\section{Field Redefinition}\label{FR}

The dual giant graviton solution has been matched to a BPS classical solution of super Yang-Mills theory on
${\mathbb R}\times$S$^3$ \cite{Hashimoto:2000zp}.
There is a non-trivial field redefinition needed when passing from the field theory to the gravitational description.
In this section we will review this field redefinition as it is needed when we compare our emergent dynamics to the
expected Yang-Mills theory.

To start, consider a Yang-Mills theory defined on ${\mathbb R}\times$S$^3$, and denote the radius of the S$^3$ by $R$.
The Abelian part of the Yang-Mills action for an adjoint scalar, after reducing to the $s$-wave, is
\bea
S={R^3\Omega_3\over 2g_{YM}^2}\int dt \left( \dot X \dot X^\dagger - {1\over R^2}X X^\dagger\,\right)
\eea
Reparametrizing the field as
\bea
X=\sqrt{g_{YM}^2 N\over R^2\Omega_3}\phi
\eea
the action becomes
\bea
S={NR \over 2}\int dt \left( \dot \phi \dot \phi^\dagger -{\phi\phi^\dagger\over R^2}\right)
\eea
Setting $\phi=\eta e^{i\omega t}$ the classical equations of motion are obeyed when
\bea
\eta =\sqrt{L\over N}
\eea
with $L$ the angular momentum of the dual giant graviton.
Further, the energy of this solution is $E=L$.
This matches the radius and energy of the dual giant graviton solution obtained using the DBI
action\cite{Grisaru:2000zn,Hashimoto:2000zp}.

This field redefinition is need for us to compare the emergent lattice dynamics to the gauge theory world volume dynamics of
the brane.
The field redefinition needed in our study has a number of interesting features.
Each row in Young diagram $r_1$ corresponds to a dual giant graviton. The number of boxes in the row gives the angular
momentum of the row and the square root of this gives the radius of the giant world volume \cite{McGreevy:2000cw}, i.e.
the $i$th giant has a radius
\bea
R=\sqrt{l_{r_i}}=\sqrt{l_{R_i}}
\eea
where the second equality is true at large $N$ in the displaced corners limit.
Next, a number of studies
\cite{deMelloKoch:2009jc,Koch:2016jnm,deMelloKoch:2018tlb,Kim:2018gwx,deMelloKoch:2018ert,Suzuki:2020oce}
have established that when fields that correspond to boxes on a large Young diagram interact,
they do so with an effective 't Hooft coupling obtained by replacing $Ng_{YM}^2\to N_{\rm eff}g_{YM}^2$, with $N_{\rm eff}$
given by the factor of the box that is interacting.
For boxes appearing in the $i$th row of $r_1$ we should replace
\bea
Ng_{YM}^2\to (N+l_{R_i})g_{YM}^2
\eea
With these two replacements, the field redefinitions needed in Section \ref{EYM} are ($a,b$ are oscillators for the $X$ and
$Y$ fields, while $b_1,b_2$ are oscillators for the $\phi_1,\phi_2$ fields)
\bea
a_{ii}=\sqrt{g_{YM}^2 (N+l_{R_i})\over l_{R_i}\Omega_3}(b_1)_{ii}
\qquad
b_{ii}=\sqrt{g_{YM}^2 (N+l_{R_i})\over l_{R_i}\Omega_3}(b_2)_{ii}
\eea
as well as the dagger of these equations.

\end{document}